\documentclass[aps,prx, showpacs, twocolumn, superscriptaddress]{revtex4-1}
\usepackage{graphicx}  % needed for figures
\usepackage{dcolumn}   % needed for some tables
\usepackage{bm}        % for math
\usepackage{amssymb}   % for math
\usepackage{amsmath}   % for math
\usepackage{bbold}
\usepackage{dsfont} % for math
\usepackage[position=top, caption=false]{subfig} % caption not compatible with revtex4-1
\usepackage{floatrow}
\usepackage{url}
\usepackage[version=3]{mhchem} % for chemistry
\usepackage{algorithm}
\usepackage[noend]{algpseudocode}
\usepackage{hhline}

\newcommand{\quotes}[1]{\lq\lq#1\rq\rq}
\newcommand{\ehpair}{$\emph{e}$-$\emph{h}$}
\newcommand{\eepair}{$\emph{e}$-$\emph{e}$}

\begin{document}
% Solution to Quantum Many-Body Problems in X-Ray Excitations Enabled by a Basic Graph Algorithm
% Solution to Quantum Many-Body Problems in X-Ray Excitations Made Possible by Breadth-First Search
% Solving Quantum Many-Body Problems in X-Ray Spectra via a Basic Graph Algorithm 
%\title{Solving Quantum Many-Body Problems in X-Ray Spectra via a Basic Graph Algorithm}
\title{Quantum Many-Body Effects in X-Ray Spectra Efficiently Computed using a Basic Graph Algorithm}

\author{Yufeng Liang}
\email{yufengliang@lbl.gov}
\affiliation{The Molecular Foundry, Lawrence Berkeley National Laboratory, Berkeley, CA 94720, USA}
\author{David Prendergast}
\affiliation{The Molecular Foundry, Lawrence Berkeley National Laboratory, Berkeley, CA 94720, USA}

\begin{abstract}
The growing interest in using x-ray spectroscopy for refined materials characterization calls for accurate electronic-structure theory to interpret x-ray near-edge fine structure.
In this work, we propose an efficient and unified framework to describe all the many-electron processes in a Fermi liquid after a sudden perturbation (such as a core hole).
This problem has been visited by the Mahan-Nozi\'eres-De Dominicis (MND) theory, but it is intractable to implement various Feynman diagrams within first-principles calculations.
Here, we adopt a non-diagrammatic approach and treat all the many-electron processes in the MND theory on an equal footing.
Starting from a recently introduced determinant formalism [Phys. Rev. Lett. 118, 096402 (2017)], 
we exploit the linear-dependence of determinants describing different final states involved in the spectral calculations.
An elementary graph algorithm, breadth-first search, can be used to quickly identify the important determinants for shaping the spectrum, which avoids the need to evaluate a great number of vanishingly small terms.
This search algorithm is performed over the tree-structure of the many-body expansion, which mimics a path-finding process.
We demonstrate that the determinantal approach is computationally inexpensive even for obtaining x-ray spectra of extended systems.
Using Kohn-Sham orbitals from two self-consistent fields (ground and core-excited state) as input for constructing the determinants, the calculated x-ray spectra for a number of transition metal oxides are in good agreement with experiments.
Many-electron aspects beyond the Bethe-Salpeter equation, as captured by this approach, are also discussed, 
such as shakeup excitations and many-body wave function overlap considered in Anderson's orthogonality catastrophe.

\end{abstract}
\maketitle

\section{INTRODUCTION}
%% applications of x-ray spectra
There is a fast-growing interest in using first-principles computational methods to interpret x-ray spectroscopies for characterizations of materials and thereby enhance our basic understanding of electronic structure \cite{haerle2001sp, nilsson2004chemical, wernet2004structure, prendergast2006x, debeer2010calibration, woicik2007ferroelectric, rehr2009ab, zhao2011visualizing, tan2012unraveling, liu2012phase, drisdell2013probing, eelbo2013adatoms, t2013compensation, velasco2014structure, pascal2014x, mcdonald2015cooperative, wernet2015orbital, lu2016quantitative, drisdell2017determining, liu2017highly}.
Fulfilling this task requires a reliable prediction of possible atomic structures that could lead to the observed spectra, and more challengingly, a generic theory that can predict accurate x-ray spectral fingerprints for given systems.
Central to a first-principles spectroscopic theory is solving the dynamics of a many-electron Hamiltonian upon excitation of a core electron by an x-ray photon, for realistic systems ranging from molecules to solids, in an efficacious manner.

From a fundamental viewpoint, the approaches to tackle a many-body problem fall into two major categories.
Quantum-field-theoretical methods \cite{wen2004quantum, bruus2004many, tsvelik2007quantum, mattuck2012guide} focus on describing the trajectories of a many-body system.
Through computing the path integrals of all trajectories from one many-body state to another,
one obtains the transition probability between the two.
The field-theoretical approach has given rise to a set of powerful first-principles tools such as the $GW$ and Bethe-Salpeter-Equation (BSE) method \cite{hybertsen1986electron, rohlfing2000electron, onida2002electronic, vinson2011bethe}.
In current implementations of these methods, only a finite set of diagrams are incorporated, due to the daunting complexity of evaluating all them. 
The other category of approaches focuses on the description of many-body wave functions based on Slater determinants \cite{knowles1984new, szabo2012modern, shao2006advances}.
This leads to methods that are used prevalently in quantum chemistry such as the full configuration interaction (FCI) approach and the coupled-cluster technique \cite{pople1987quadratic, bartlett2007coupled, booth2013towards},
or exact diagonalization for solving strongly-correlated systems \cite{caffarel1994exact, dagotto1994correlated}.
Currently these methods are mostly applied to systems with $10-20$ electrons limited by the exponential growth of the configuration space. 

For x-ray excitations and associated spectra, we have witnessed the success of the constrained-occupancy density functional theory ($\Delta$SCF) 
\cite{taillefumier2002x, prendergast2006x, drisdell2013probing, pascal2014x, velasco2014structure, ostrom2015probing, drisdell2017determining}, 
which approximates an x-ray excited state with one empty Kohn-Sham (KS) orbital in the final state. 
Recently, we highlighted the shortcomings in this single-particle (1p) approach for a class of $3d$ transition metal oxides (TMOs) 
and indicated an lack of generation to higher-order excitations involving multiple electron-hole (\ehpair{}) pairs \cite{liang2017accurate}. 
Driven by these deficiencies, we proposed a better many-body wavefunction ansatz
that approximates the initial and final state with a \emph{single} Slater determinant.
The initial-state Slater determinant is constructed from the KS orbitals of the ground-state system, 
while the Slater determinant for a specific final-state is derived from the KS orbitals of the core-excited system. 
Within this approximation,
the transition amplitude can also be expressed as a determinant \cite{anderson1967infrared, dow1980solution, stern1983many, ohtaka1990theory, liang2017accurate} comprising transformation coefficients between the two KS basis sets.
We find this determinant approach can rectify the deficiency of the 1p $\Delta$SCF approach for a few TMOs \cite{liang2017accurate}.
It is natural to ask:
(a) does this formalism provide a good approximation for x-ray near-edge structures in general?
(b) is it practicable for calculations of extended systems, given the huge configuration space?
(c) can this approach permit access to higher-order excitations and describe various many-body x-ray spectral features beyond the BSE?

In this work, we answer these questions by demonstrating an efficient yet simple approach to explore the large configuration space in the determinant formalism.
A crucial first step is to relate similar determinants to one another via exterior algebra \cite{yokonuma1992tensor, winitzki2010linear}
and then evaluate them via updates, rather than from scratch.
Even so there are still $10^6$ to $10^9$ many-body states to consider for configurations with double \ehpair{} pairs .
However, only a small portion of these determinants have significant transition amplitudes, due to the spatially localized nature of core-level excitations,
as can be tested by brute-force calculations.
Motivated by this observation, we adopt a breadth-first search (BFS) algorithm \cite{knuth1998art, cormen2009introduction} to look for nontrivial configurations rather than exhausting the entire configuration space.

The BFS algorithm is a basic algorithm for traversing a tree structure, finding the shortest path \cite{skiena1990dijkstra, zeng2009finding}, solving a maze \cite{moore1959shortest}, and other combinatorial search problems.
Although the BFS algorithm cannot guarantee answers within a polynomial time, substantial speed-up can often be achieved via heuristically pruning the search tree \cite{pearl1984heuristics, zhou2006breadth}.
For the many-body configuration problem, we design the BFS to search for active \lq\lq pathways\rq\rq{} from the initial state to many excited-state configurations.
Instead of directly accessing a large number of high-order configurations, 
the search algorithm first visits its ascendant configurations with fewer \ehpair{} pairs.
If multiple pathways to an ascendant configuration interfere destructively and result in a small transition amplitude,
the search algorithm will discard the configuration before more high-order configurations are generated. 
We will show that this tree-pruning technique can typically lead to at least 100-fold speed-up in the calculation of x-ray spectra.
This search algorithm is generic and can be generalized to any kind of sudden perturbation.

The determinant formalism is an exact solution to the Mahan-Nozi\'eres-De Dominicis (MND) model \cite{mahan1967excitons, nozieres1969singularities} in which multiple electrons interact with a core hole.
Hence, this approach can naturally incorporate all many-electron processes in the MND theory, 
which includes 
the direct and exchange diagrams as in the BSE \cite{rohlfing2000electron, onida2002electronic, vinson2011bethe},
the zig-zag diagrams,
and the diagrams with a core hole dressed by many \ehpair{} bubbles.
While the BSE diagrams mainly describe \ehpair{} attraction, or excitonic effects,  
the zigzag or bubble diagrams describe higher order \ehpair{} excitations 
that lead to shakeup features \cite{brisk1975shake, mcintyre1975x, ohtaka1990theory, enkvist1993c, calandra2012k, mahan2013many, lemell2015real}
or many-body effects due to reduced wave-function overlap.
A reduction in many-body wave function overlap is the origin of the Anderson orthogonality catastrophe \cite{anderson1967infrared, mahan2013many}.
If one were to include all of these effects using the diagrammatic approach,  a comprehensive set of techniques, such as solving BSE-like equations and using a cumulant expansion \cite{kas2015real, kas2016particle}, would be required.
Here, the determinant formalism, in conjunction with the first-principles KS orbitals, provides a efficient means to investigate all many-electron effects within the MND model rigorously, for a wide energy range, within a simple unified framework.

This new determinant formalism has already shown great practicality to address realistic problems in materials characterization. 
We systematically study the \ce{O} $K$-edge ($1s\rightarrow np$) x-ray absorption spectra (XAS) of various TMOs
and find this approach can faithfully reproduce the experimental x-ray line shapes for most of the investigated systems.
This can be immediately applied to study various energy conversion and storage systems involving oxides \cite{yabuuchi2011detailed, hu2013origin, suntivich2011design, lin2016metal, luo2016charge, strasser2010lattice, matsukawa2014enhancing, lebens2016direct, de2016mapping}, 
where the interpretation of x-ray spectra can be challenging, and the conclusions often depend sensitively on intricate near-edge line shapes.

This work is organized as follows.
Sec. \ref{sec:model} revisits the many-body effects captured by the MND theory in terms of Feynman diagrams.
Sec. \ref{sec:det} and \ref{sec:comparison}  provides a solution to the MND model from the perspective of many-electron wave functions and introduces the determinant formalism.
Sec. \ref{sec:sol} introduces exterior algebra to elucidate the linear dependence of the determinants that is encoded in the so-called $\zeta$-matrix, 
followed by a BFS algorithm for an efficient evaluation in Sec. \ref{sec:bfs}.
Sec. \ref{sec:dscf} discusses how to combine this algorithm with DFT simulations and its validity in the presence of \eepair{} interactions.
The simulated XAS of a variety of oxides are shown in Sec. \ref{SEC:results}, 
with analysis of spectra obtained from different level of approximations.
The many-body aspects beyond the BSE as captured by this method will be discussed in Sec. \ref{sec:more_eh} and \ref{sec:spin_conv}, using the half-metal \ce{CrO2} as an example.
Finally, the numerical details and efficiency of the newly introduced algorithm are analyzed in Sec. \ref{sec:comp}.

\section{THEORETICAL MODELS AND METHODS}

\subsection{Independent-electron model and diagrammatic approaches}
\label{sec:model}
We first revisit the conceptually simple MND model from the perspective of Feynman diagrams.
The incorporation of first-principles calculations will be deferred to Sec. \ref{sec:dscf}.
In the MND model \cite{nozieres1969singularities, ohtaka1990theory, mahan2013many}, the electrons only interact with the core hole and electron-electron (\eepair{}) interactions are neglected.
Consider a supercell with one of the atoms replaced by its core-excited version. 
This is typically a good approximation to a core-excited system at low photon flux.
Assume there are $N$ valence electrons in its ground state and there is only one core level. 
The MND Hamiltonian without \eepair{} interactions reads
\begin{align}
\begin{split}
\mathcal{H}&=\mathcal{H}_0+\mathcal{H}_I\\
\mathcal{H}_0&=
\sum_{c}\varepsilon_{c} a^\dagger_{c} a_{c} 
- \varepsilon_h h^\dagger h\\
\mathcal{H}_I&=
\sum_{cc'} V_{cc'} a^\dagger_{c} a_{c'}
h^\dagger h\\
\end{split}
\label{eq:ham_mnd}
\end{align}
where the diagonal part $\mathcal{H}_0$ is composed of the valence orbitals ($c$ iterates over both occupied and empty valence orbitals) and the core level ($h$).
$a^\dagger_c$ and $h^\dagger$ are electron and hole creation operators respectively.
The only two-body term in $\mathcal{H}$ is the Coulomb interaction between the valence orbitals and the core level, as described by $\mathcal{H}_I$, 
in which the core-hole potential $V_{\alpha\beta}$ is defined by
\begin{align}
V_{\alpha\beta}&=\int d^3 r d^3 r' \psi^*_\alpha(\textbf{r})\psi_\beta(\textbf{r})V(\textbf{r}, \textbf{r}')
\psi^*_h(\textbf{r}')\psi_h(\textbf{r}')
\label{eq:v}
\end{align}
where $\psi_i$'s are the 1p wave functions and $V(\textbf{r}, \textbf{r}')$ is the (effective) Coulomb potential.
The two-body interaction $V_{\alpha\beta}$ accounts for the electron scattering from orbital $\beta$ to $\alpha$ due to the core-hole potential.

The x-ray photon field can be described by a current operator \cite{mahan2013many} that promotes one core electron to a valence orbtial
\begin{align}
\begin{split}
\hat{J}=\sum_{c} a^\dagger_c h^\dagger\langle\psi_c|\hat{j}|\psi_h\rangle + h.c. 
\label{eq:J}
\end{split}
\end{align}
The transition operator is the electric field polarization-projected position operator that couples the core level to valence orbitals: $\hat{j} = \bm{\epsilon}\cdot\bm{r}$, in the limit of zero-momentum transfer and within the dipole approximation \cite{prendergast2006x, de2008core}.
In principle, the transition operator $\hat{j}$ can be any other local sudden perturbation, not necessarily limited to a core hole.

The independent-electron model was originally considered by the MND theory \cite{mahan1967excitons, nozieres1969singularities, mahan2013many} using diagrammatic techniques.
The time-evolution of the many-electron system after photon absorption is described by the Kubo current-current correlation function
\begin{align}
\begin{split}
\Pi(t) &= -\frac{i}{\mathcal{V}} \langle  \Psi_i| \mathcal{T}  \hat{J}(t)  \hat{J}(0)]  \Psi_i\rangle\\
&=\frac{1}{\mathcal{V}}\sum_{cc'} \langle\psi_c|\hat{j}|\psi_h\rangle  \langle\psi_h|\hat{j}|\psi_{c'}  \rangle L_{cc'}(t)\\
&=\frac{1}{\mathcal{V}} \sum_{cc'} w_c w^*_{c'} L_{cc'}(t)
\end{split}
\label{eq:pi_t}
\end{align}
where $w_c = \langle\psi_c|\hat{j}|\psi_h\rangle$ is the vertex that represents the absorption of a photon to create an \ehpair{} pair  ($w^*_c$ represents the opposite process).
The x-ray absorption spectrum (XAS) $A(\omega)$ is the spectral function of the photon self-energy in the frequency domain
\begin{align}
\begin{split}
\Pi(\omega) &= \int^\infty_{-\infty} dt e^{i\omega t} \Pi(t)\\
A(\omega) & = -\frac{1}{\pi}\text{Im} \Pi(\omega)
\end{split}
\end{align}
In the following discussion, we focus on the \ehpair{} correlation function as defined in Eq. (\ref{eq:pi_t})
\begin{align}
\begin{split}
L_{cc'}(t) = -i  \langle \Psi_i|\mathcal{T} h(t) a_c(t) a^\dagger_c(0) h^\dagger(0) |\Psi_i\rangle
\end{split}
\end{align}
which includes all the many-electron processes in x-ray absorption.

\begin{figure}[t]
  \centering
  \includegraphics[angle=0, width=0.90\linewidth]{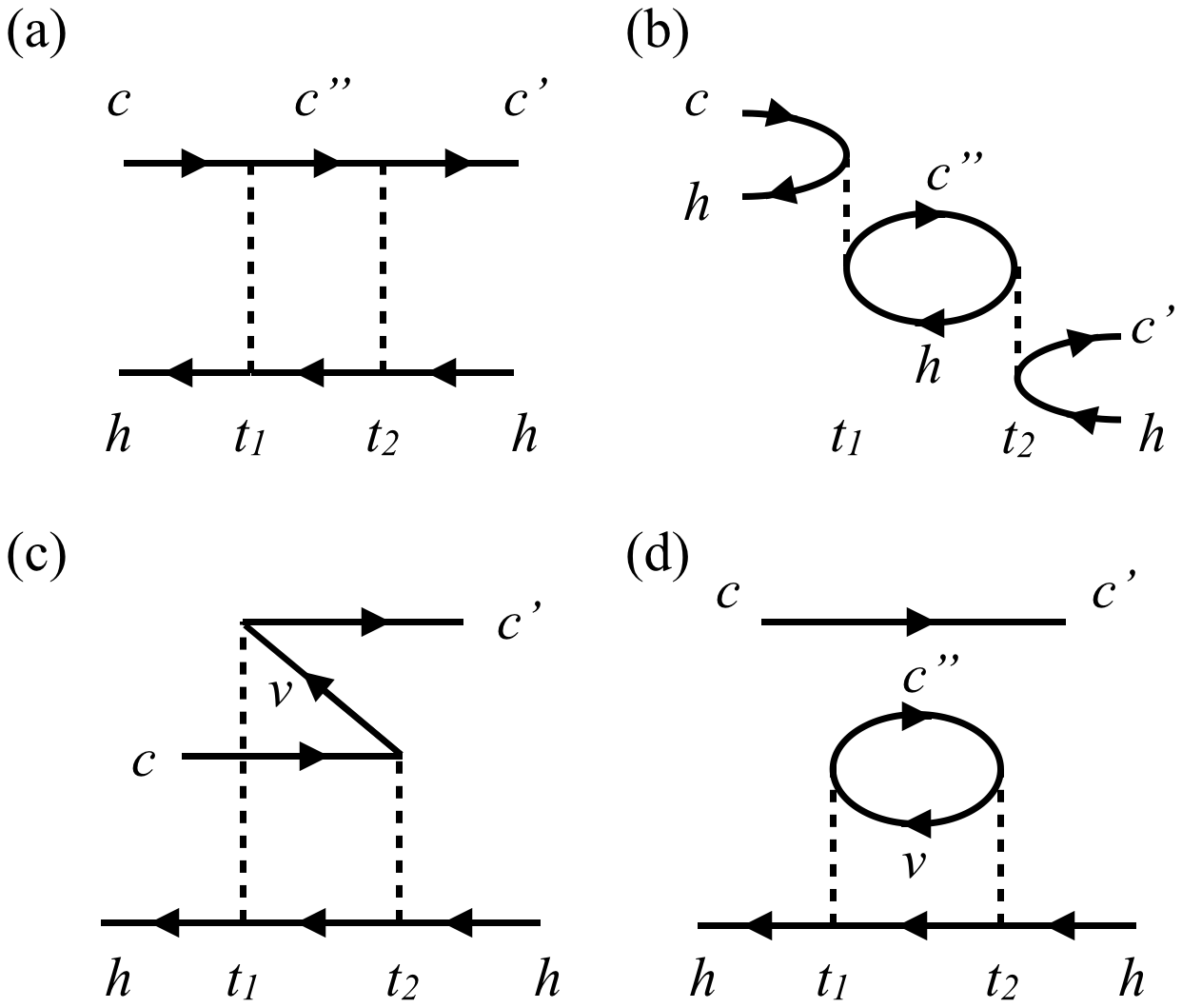} % put t_1 and t_2 on; c'' --> c' 
  \caption{
  Four distinct types of \ehpair{} processes in the second-order Feynman diagrams in the MND theory. 
  There are exactly two Coulomb lines (at $t_1$ and $t_2$) in each diagram, as marked by vertical dashed lines.
  }
  \label{fig:lcc_2nd}
\end{figure}

We exemplify these many-electron processes by four types of second-order Feynman diagram of $L_{cc'}(t)$, as shown in Fig. \ref{fig:lcc_2nd}. 
The time axis runs from left to right and the Coulomb lines are vertical due to the neglect of dynamical effects in the Coulomb interaction $\mathcal{H}_I$.
The BSE captures two kinds of processes: direct \ehpair{} attraction as described by the ladder diagram Fig. \ref{fig:lcc_2nd} (a), and \ehpair{} exchange as described by the diagram in Fig. \ref{fig:lcc_2nd} (b).
In these diagrams, there is only one \ehpair{} pair present at any time of the propagation. 
However, there are other diagrams with more \ehpair{} pairs present at a time, e.g., the zigzag diagram in Fig. \ref{fig:lcc_2nd} (c).
The corresponding process involves a core hole causing the ground state to decay into a valence \ehpair{} pair ($c'$ and $v$) at $t_1$.
At a later time $t_2$, the core hole assists the newly generated valence hole ($v$) to recombine with incoming electron ($c$), leaving an outgoing electron ($c'$) and the core hole.
Lastly, it is also possible that the valence \ehpair{} pair ($c'$ and $v$) generated earlier does not correlate with the incoming electron at all and simply annihilates at a later time $t_2$. 
This leads to a bubble diagram with a freely propagating electron and a core hole dressed by \ehpair{} bubbles as shown in Fig. \ref{fig:lcc_2nd} (d). 
These \ehpair{} bubbles tend to reduce the many-body wave function overlap and are the causes for the Anderson orthogonality catastrophe \cite{anderson1967infrared}.

The MND theory \cite{mahan1967excitons, nozieres1969singularities, mahan2013many} systematically studies and estimates the impact of these diagrams on the near-edge structure of x-ray spectra. 
In essence, it is found that denominators in the BSE diagrams involve $\varepsilon_{c} - \varepsilon_{h}$, which is roughly the energy required to create an electron-core-hole excitation, while the denominators in the zigzag or bubble diagrams involve an offset of $\varepsilon_c - \varepsilon_v$, the energy required to create an additional (valence) \ehpair{} pair.
This means the zigzag processes or the bubble diagrams can become significant in a metallic system where $\varepsilon_c - \varepsilon_v$ can be vanishingly small, or if the photon energy is sufficiently high to be in resonance with double \ehpair{} excitations.

In practical first-principles BSE calculations however, \eepair{} interactions are taken into account 
and the bare Coulomb interactions in the direct (Fig. \ref{fig:lcc_2nd} (a) ) 
diagram are replaced by screened Coulomb interactions.
The screened Coulomb interactions are typically modeled with the empty-bubble diagrams within the random-phase approximation \cite{rohlfing2000electron, onida2002electronic, vinson2011bethe}, 
which, to some extent, describe the many-electron screening effects in x-ray excitations.

\subsection{An alternative MND solution based on many-body wave functions}
\label{sec:det}
In the last section we have discussed the diagrammatic approach, or many-body perturbation theory (MBPT),  for solving the MND model  [Eq. (\ref{eq:ham_mnd})].
However, this Hamiltonian is essentially quadratic and exactly solvable. 
%The sudden approximation assumes that the core-hole is a long-lived quasiparticle  with respect to the electronic response to that excitation, ignoring, for subsequent spontaneous decay of the core-hole excited state, via Auger emission, for example. Therefore, assuming that the core-holes define stationary states, i.e., with no quantum fluctuation in the number of core holes, the Hamiltonian can be converted into a quadratic form by taking the expectation value of $h^\dagger h$.
For the initial state, no core hole is excited and $\langle h^\dagger h \rangle = 0$ and hence the initial-state Hamiltonian $\mathcal{H}_i$ is simply $\mathcal{H}_0$
For the final state, there is exactly one core hole, i.e., $\langle h^\dagger h \rangle = 1$, and the final-state Hamiltonian $\mathcal{H}_f$ also becomes quadratic
\begin{align}
\begin{split}
\mathcal{H}_f& = \mathcal{H}_i  + \sum_{cc'}V_{cc'} a^\dagger_{c} a_{c'}
\end{split}
\label{eq:hi_hf}
\end{align}
%We will assume, for the moment, that both the initial- and final-state Hamiltonians have solutions and delay discussion of practical approximations to Sec.~\ref{sec:dscf}. 
Within the quadratic forms, it is straightforward to construct the many-body wave functions of the initial- and final-state.
The initial state is simply a Slater determinant that consists of $N$ valence electrons occupying the $N$ lowest-lying orbitals:
%These orbitals and their energies can be obtained from DFT calculations or its derivates, as will be shown in Sec. \ref{sec:results}.
%When expressed in second quantization, the many-body wave function reads
\begin{align}
\begin{split}
|\Psi_i\rangle = \big( \prod^N_{\mu=1} a^\dagger_\mu \big) h |0\rangle
\end{split}
\label{eq:psi_i}
\end{align}
where $\mu$ goes over all the occupied valence orbitals, $h$ annihilates the core hole (fills the core level with one electron), and $|0\rangle$ is the null state with no electrons.
The final-state XAS wave functions can be expressed in a similar manner, but using
the eigenvectors of $\mathcal{H}_f$
\begin{align}
\begin{split}
|\Psi_f\rangle = \prod^{N+1}_{\mu=1} \tilde{a}^\dagger_{f_{\mu}}  |0\rangle
\end{split}
\label{eq:psi_f0}
\end{align}
where the index $f$ is a tuple: $f=(f_1, f_2, \dots, f_{N+1})$, which denotes the \emph{valence} $N+1$ orbitals that the $N+1$ electrons will occupy in the final state.
$\tilde{a}_i$ (with tilde) correspond to the eigenvectors of $\mathcal{H}_f $ so that $\mathcal{H}_f = \sum_i \tilde{\varepsilon}_i \tilde{a}^\dagger_i \tilde{a}_i$.
To apply the Fermi's Golden rule, one needs to work within the same basis set.
We express the final-state basis set in terms of the initial-state one:
\begin{align}
\begin{split}
|\tilde{\psi}_i \rangle &= \sum_j \xi_{ij} | \psi_j \rangle\\
\tilde{a}^\dagger_i &= \sum_j \xi_{ij} a^\dagger_j
\end{split}
\label{eq:xi}
\end{align}
where $\xi_{ij}$'s are the transformation coefficients: $\xi_{ij} = \langle  \psi_j |  \tilde{\psi}_i \rangle$.

With these expressions for $|\Psi_i\rangle$ and $|\Psi_f\rangle$, the many-body transition matrix element for any one-body operator $\mathcal{O}$ has been calculated in previous work \cite{stern1983many, ohtaka1990theory, liang2017accurate}
\begin{align}
\begin{split}
\langle \Psi_f | \mathcal{O} | \Psi_i \rangle = \sum_{c} (A^{f}_c)^* \langle \psi_c | o | \psi_h \rangle 
\end{split}
\label{eq:mat_elem}
\end{align}
in which the transition amplitude also takes a determinantal form
\begin{align}
\begin{split}
A^{f}_c=
\det
\begin{bmatrix}
\xi_{f_1, 1} & \xi_{f_1, 2} & \cdots & \xi_{f_1, N} & \xi_{f_1, c} \\
\xi_{f_2, 1} & \xi_{f_2, 2} & \cdots & \xi_{f_2, N} & \xi_{f_2, c} \\
\vdots & & \ddots & & \vdots\\
\xi_{f_{N+1}, 1} & \xi_{f_{N+1}, 2} & \cdots & \xi_{f_{N+1}, N} & \xi_{f_{N+1}, c} \\
\end{bmatrix}
\end{split}
\label{eq:afc}
\end{align}
The row index goes over $N+1$ occupied final-state orbitals $f_i$, 
the column index over the lowest-lying $N$ initial-state orbitals plus one \emph{empty} orbital labeled by $c$ 
(This empty orbital is coupled to the core level with the one-body operator $\mathcal{O}$).
This determinantal form reflects how these $N+1$ electrons transit from the initial to final state in the x-ray excitation process.
All the possible electronic pathways are taken into account by the transformation matrix in $A^f_c$.
The transition amplitude of an individual electron is quantified by the matrix elements, i.e., the initial-final orbital overlap $\xi_{ij} = \langle  \psi_j |  \tilde{\psi}_i \rangle$.
The interference of these pathways is lumped into a determinant due to the fermionic nature of electrons.

For the quadratic $\mathcal{H}_f$, the energy of a final-state $|\Psi_f\rangle$ can be obtained 
by direct summation of 1p-orbital energies
\begin{align}
\begin{split}
E_f=\sum^{N+1}_{j=1}\tilde{\varepsilon}_{f_j}
\label{eq:Ef}
\end{split}
\end{align}
where $\tilde{\varepsilon}_{f_j}$ are taken from the diagonalized $\mathcal{H}_f$.
A relative energy $\Omega_f = E_f - E_\text{th}$ may also be defined for later discussion,
where $E_\text{th}$ is the energy of the lowest-lying $|\Psi_f\rangle$:
$E_\text{th}=\sum^{N+1}_{j=1}\tilde{\varepsilon}_{j} $.

\begin{figure}[t]
  \centering
  \includegraphics[angle=0, width=0.90\linewidth]{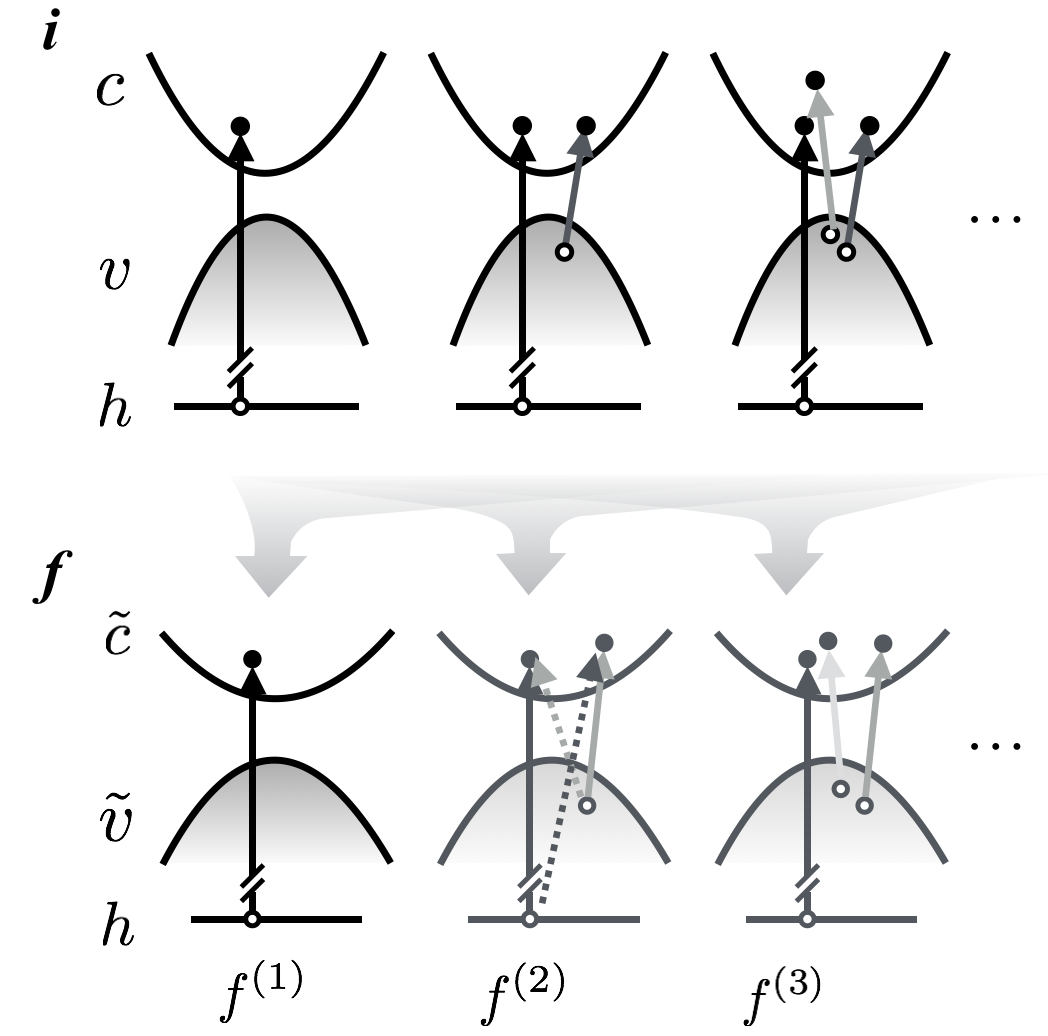} 
  \caption{
Definitions of the multi-electron configurations used in the initial(i)- and final(f)-state picture according to the convention in quantum chemistry. 
A final-state configuration (a single Slater determinant) at the order of $f^{(n)}$ can be hybridized from a number of initial-state configurations at multiple orders, 
as shown by the thick opaque downarrows, which illustrates the spirit of Eq. (\ref{eq:psi_f0}).
The solid uparrows in a configuration indicate one possible multi-electronic pathway to access that configuration from the ground-state.
The dashed uparrows show the other possible pathway to access the $f^{(2)}$ configuration.
  }
  \label{fig:fn}
\end{figure}

For ease of calculation, previously we have also regrouped the final-state multi-electron configurations according to the convention in quantum chemistry \cite{dow1980solution, szabo2012modern, bartlett2007coupled}.
The configuration $f = (1, 2, \cdots, N, c_0)$ with $c_0 > N$ is dubbed as a \emph{single} or a $f^{(1)}$ configuration 
because it has one electron-(core-)hole pair.
A shorthand notation for an $f^{(1)}$ configuration can be employed, using $(c)$ to denote the the orbital of the excited valence electron.
$f=(1,2,\cdots, v_1 - 1, v_1 + 1, \cdots, N, c_0, c_1)$ with $v_1 \leq N$ and $c_1 > c$ is dubbed as a \emph{double} or $f^{(2)}$ configuration because it has one extra (valence) \ehpair{} pair
as defined by the electron (hole) index $c_1$ ($v_1$). 
The shorthand notation for $f^{(2)}$ is $(c_0, v_1, c_1)$.
This definition can be extended to higher orders such as triples and so forth.
For unique indexing, we require $c_0 < c_1 < c_2 < \cdots < c_{n - 1}$ and $v_1 > v_2 > \cdots > v_{n - 1}$ in a $f^{(n)}$ index.
Examples of final-state $f^{(n)}$ are shown in Fig. \ref{fig:fn} (schematics on the second row).

\subsection{Interpretation of the final-state many-body approach from an initial-state perspective}
\label{sec:comparison}

In this section, we provide a comparison between the outlined determinant formalism and MBPT using Feynman diagrams.
While the determinant formalism constructs many-electron states using both initial- and final-state orbitals, MBPT, such as BSE, relies on initial-state quantities only. 
To relate the two theories, we can express the MND many-electron final states
 $| \Psi_f \rangle$ in Eq. (\ref{eq:psi_f0}) using only the initial-state orbitals.
Rewriting final-state operators $\tilde{a}_i$ according to a linear combination of the initial-state operators $a_i$ [Eq. (\ref{eq:xi})] and expressing the wave function $|\Psi_f\rangle$ in terms of $|\Psi_i\rangle$:
\begin{align}
\begin{split}
|\Psi_f\rangle &= \prod^{N+1}_{\mu=1} \sum_{j_\mu} \xi_{f_\mu, j_\mu} a^\dagger_{j_\mu}  |0\rangle \\
 &= \prod^{N+1}_{\mu=1} ( \sum_{j_\mu} \xi_{f_\mu, j_\mu} a^\dagger_{j_\mu} ) \big( \prod^N_{\nu=1} a_\nu \big) h^\dagger |\Psi_i\rangle
\end{split}
\label{eq:psi_f_all}
\end{align}
Expanding the product of the operators and regrouping like terms,
\begin{align}
\begin{split}
|\Psi_f\rangle 
&= \sum_{c\in\text{unocc}} A^f_c a^\dagger_c h^\dagger |\Psi_i\rangle \\
&+ \sum_{\substack{c, c'\in\text{unocc} \\ v \in \text{occ}}} B^f_{cc',v} (a^\dagger_c h^\dagger) (a^\dagger_{c'} a_v) |\Psi_i\rangle \\
&+ \cdots
\end{split}
\label{eq:psi_f}
\end{align}
The leading-order term comprises linear combinations of single electron-(core-)hole pairs, because there are $N+1$ creation operators $a_i^\dagger$ and $N$ destruction operators $a_i$ in Eq. (\ref{eq:psi_f_all}),
leaving at least one creation operator $a^\dagger_c$ for an unoccupied state.
For this term, $N$ out of $N+1$ indices $j_\mu$ are chosen from $1,2,\cdots, N$ so that $N$ $a^\dagger_i$'s can cancel with $N$ $a_i$'s.
There are $(N+1)!$ such permutations, and reordering the fermionic operators gives rise to the determinantal form of the coefficients, as previously stated in Eq. (\ref{eq:afc}). 

The next term in Eq. (\ref{eq:psi_f}) is a double term $(a^\dagger_c h^\dagger) (a^\dagger_{c'} a_v)$, 
which has one additional valence \emph{e-h} pair $(a^\dagger_{c'} a_v)$ generated on the top of the electron-core-hole pair.
This term takes into account the second-order many-electron processes: the valence \emph{e-h} excitations induced by the core-hole potential, which are also known as the \emph{shakeup} excitations \cite{brisk1975shake, mcintyre1975x, ohtaka1990theory, enkvist1993c, calandra2012k, mahan2013many, lemell2015real},
because an additional amount of energy is required to create these valence excitations. % which depends on the energy difference $\varepsilon_c - \varepsilon_{v'}$.
As the series expansion proceeds, each term will have one more valence $\emph{e-h}$-pair than the last, 
and more complicated shakeup processes with multiple \emph{e-h}-pairs are included.
A full schematic for the relation of one single final-state configuration $|\Psi_f\rangle$ (written as one Slater determinant using final-state orbitals) in terms of initial-state configurations is  shown in Fig. \ref{fig:fn}.

Within MBPT, the configuration series in Eq. (\ref{eq:psi_f}) is typically truncated,
and the coefficients are solved by expanding the Hamiltonian over the restricted configuration space and solving the eigenvalue problem.
In the BSE, for instance, the final-state Hamiltonian is expanded over the single-\emph{e-h}-pair space $a^\dagger_c h^\dagger |\Psi_i\rangle$
and the eigenvector coefficients (analogous to $A^f_c$) refer to this single-\ehpair{} basis.
In some sense, this approximation corresponds to the ladder and exchange diagrams: at any point in time of the propagation, there is only one \emph{e-h} pair involved.

By contrast, the determinant formalism does not restrict the number of \emph{e-h}-pairs in the final-state configuration space.
When $|\Psi_f\rangle$ is projected onto $|\Psi_i\rangle$ as in Eq. (\ref{eq:psi_f}), a superposition of single, double, and high-order terms naturally arises,
although only the leading-order coefficients $A^f_c$ are relevant for calculating matrix elements of one-body operator.
In this way, the zig-zag and bubble diagrams, present within MND theory, which involve multiple \emph{e-h}-pair generation, are automatically incorporated.

\subsection{Efficient evaluation of determinantal transition amplitudes}
\label{sec:sol}
The above determinantal formalism provides an alternative solution to the MND model in Eq. (\ref{eq:ham_mnd})
without using diagrammatic approaches. 
If a sufficient number of final states are included, one may expect the determinantal method to give the spectrum as solved from the MND model.
However, an brute-force calculation is rarely used because the many-electron configuration space grows factorially with the number of electrons. 
It does not seem to be practical to compute the large number of determinants that would represent all configurations.

For a half-filled system with $M$ orbitals and $N$ ($N \approx M / 2$) electrons, even the $f^{(2)}$ group has $(M - N) (M - N - 1) N \approx M^3 / 8$ configurations.
Iterating the index $c$ of $A^f_c$ [Eq. (\ref{eq:afc})] over all empty initial-state orbitals multiplies the time complexity by a factor of $M / 2$. 
Calculating the determinant for each configuration requires a computational cost of $\mathcal{O}(M^3)$. 
With all the three factors combined, obtaining the determinants for all of the $f^{(2)}$ configurations gives rise to a time complexity of $\mathcal{O}(M^7)$.
For metallic systems where the fermi surfaces are susceptible to the core-hole potential, higher-order terms such as  $f^{(3)}$ are typically needed for testing convergence, which leads to a higher time complexity of $\mathcal{O}(M^9)$.
Such a brute-force calculation that scales up quickly with number of states is not very practical for realistic core-hole calculations in which there could easily be $10^2$ to $10^3$ orbitals.

In this section, we introduce an efficient algorithm at much lower computational cost to access the determinants that are important for determining the x-ray spectrum.
The $\mathcal{O}(M^3)$ determinant calculation needs to be performed only once for a given configuration, and subsequently the determinants for other configurations can be derived from it.
More importantly, a BFS algorithm is employed to identify the important determinants above a specified threshold, largely reducing the number of configurations to be visited.

An apparent first step is to move the summation over $c$ in Eq. (\ref{eq:mat_elem}) into the definition of the transition amplitude coefficient, so that for each final-state configuration $f$, obtaining $A^f = \langle \Psi_f |\mathcal{O}| \Psi_i \rangle$ requires calculating only one determinant.
More specifically, we rewrite $A^f$ as
\begin{align}
\begin{split}
A^f & = \det \bm{A}^f\\
\bm{A}^f & =  
\begin{bmatrix}
\xi_{f_1, 1} & \xi_{f_1, 2} & \cdots & \xi_{f_1, N} & \sum_c \xi_{f_1, c} w^*_c\\
\xi_{f_2, 1} & \xi_{f_2, 2} & \cdots & \xi_{f_2, N} & \sum_c \xi_{f_2, c} w^*_c \\
\vdots & & \ddots & & \vdots\\
\xi_{f_{N+1}, 1} & \xi_{f_{N+1}, 2} & \cdots & \xi_{f_{N+1}, N} & \sum_c \xi_{f_{N+1}, c} w^*_c \\
\end{bmatrix}
\end{split}
\label{eq:Af}
\end{align}
where $w_c = \langle \psi_c |o| \psi_h \rangle$. 
The summation in the $(N+1)^{th}$ column of $\bm{A}^f$ can be calculated first before obtaining the determinant.
This reduces the overall time complexity by a factor of $M$.

Secondly, when considering transitions to various final-state configurations, the determinants of interest in fact have many common rows
so one can make use of the multilinearity of determinants to speed up the calculations significantly.
For example, the tuple for a double configuration $(1,2,\cdots, v_1 - 1, v_1 + 1, \cdots, N, c, c_1)$ only differs from the ground-state one $(1,2,\cdots, N, N + 1)$ by 3 indices, 
meaning their corresponding determinants $A^f$ only differ by 3 rows.
This observation motivates us to choose the determinant for the ground state as a reference, and evaluate other determinants for excited states via 
a low-rank updating technique.
%a common technique that has been used in quantum Monte Carlo (QMC).

To demonstrate this technique, it is most transparent to express the determinant in terms of the wedge (exterior) product \cite{yokonuma1992tensor, winitzki2010linear} of its row/column vectors.
The wedge product is anticommutative and has similar algebra to the Fermionic operators.
Suppose an arbitrary matrix $\bm{A}$ has $n$ row/column vectors $a_1, a_2, \cdots, a_n$, its determinant can be expressed as
\begin{align}
\begin{split}
\det \bm{A} = a_1 \wedge a_2 \wedge \cdots \wedge a_n
\end{split}
\end{align}
Assume $\det \bm{A}$ has been calculated from scratch and is nonzero (assume $\bm{A}$ is full-rank).
If $a_n$ is replaced by a new vector $a_{n+1}$, 
which can be considered as a rank-1 update,
the updated determinant can be obtained by expanding $a_{n+1}$ in terms of $a_1, a_2, \cdots, a_n$
\begin{align}
\begin{split}
\det \bm{A}' 
&\equiv a_1 \wedge a_2 \wedge \cdots \wedge a_{n+1} \\
& = a_1 \wedge a_2 \wedge \cdots \wedge \sum_{i=1}^{n} \zeta_{n+1, i} a_i \\
& = \zeta_{n + 1, n} a_1 \wedge a_2 \wedge \cdots \wedge a_n \\
& = \zeta_{n + 1, n} \det \bm{A}
\end{split}
\label{eq:Ap}
\end{align}
where $\zeta_{ij}$ is the expansion coefficient defined as
\begin{align}
\begin{split}
a_{n+1} = \sum_{i=1}^{n} \zeta_{n+1, i} a_i
\end{split}
\label{eq:zeta_intro}
\end{align}
$\zeta_{n + 1, i}$ can be obtained via the matrix inversion of $\bm{A}$: $\zeta_{n + 1, i} = \sum_{j} a_{n + 1, j}(\bm{A}^{-1})_{ji}$.
When multiplied by $a_1 \wedge a_2 \cdots \wedge a_{n-1}$, only $a_n$ survives in the summation because $a_i \wedge a_i = 0$.
Then the new determinant $\det \bm{A}'$ is simply the product of an expansion coefficient $\zeta_{n+1, n}$ and the already-known $\det \bm{A}$.

Now if the last two lines of $\bm{A}$ are replaced by two new row vectors $a_{n+1}$ and $a_{n+2}$,
the rank-2 updated determinant is
\begin{align}
\begin{split}
&\det \bm{A}'' \\
\equiv & a_1\wedge \cdots \wedge a_{n - 2} \wedge a_{n+1} \wedge a_{n+2} \\
= & a_1 \wedge \cdots \wedge a_{n - 2} \wedge \sum_{i=1}^{n} \zeta_{n+1, i} a_i \wedge \sum_{j=1}^{n} \zeta_{n+2, j} a_j  \\
= & a_1 \wedge \cdots \wedge a_{n - 2} \wedge ( \zeta_{n+1, n-1} \zeta_{n+2, n} a_{n-1} \wedge  a_n \\
+ & \zeta_{n+1, n} \zeta_{n+2, n-1} a_{n} \wedge  a_{n-1} ) \\
= & a_1 \wedge \cdots \wedge a_{n - 2} \wedge ( \zeta_{n+1, n-1} \zeta_{n+2, n} a_{n-1} \wedge  a_n \\
- & \zeta_{n+1, n} \zeta_{n+2, n-1} a_{n-1} \wedge  a_n ) \\
= &\det 
\begin{bmatrix}
\zeta_{n+1, n-1} & \zeta_{n+1, n} \\
\zeta_{n+2, n-1} &  \zeta_{n+2, n} \\
\end{bmatrix}
\det \bm{A}
\end{split}
\label{eq:App}
\end{align}
The minus sign arises from the anticommutative property of the wedge product: $a_i \wedge a_j = - a_j \wedge a_i$.
Thus the new determinant is the product of a $2\times2$ determinant composed of the expansion coefficients and $\det A$.
The above procedure can be carried out to more general situations where more row/column vectors are replaced.
This remove the need to calculate the new determinant from scratch using the $\mathcal{O}(n^3)$ algorithm.
For a rank-$r$ update, one only needs to compute the product of the reference determinant $A^\text{ref}\equiv\det\bm{A}$ and a small $r\times r$ determinant containing $\zeta_{ij}$, 
at the cost of $\mathcal{O}(1)$.

In the context of the determinantal formalism as in Eq. (\ref{eq:Af}), 
we define the row vector corresponding to the $i^{th}$ final-state orbital as: 
\begin{align}
\begin{split}
a_i = 
\begin{bmatrix}
\xi_{i, 1} & \cdots & \xi_{i, N} &  \sum_c \xi_{i, c} w^*_c
\end{bmatrix}
\end{split}
\end{align}
Then the ground-state reference determinant can be expressed as $A^\text{ref} = a_1 \wedge a_2 \wedge \cdots \wedge a_{N} \wedge a_{N + 1}$.
To access the determinants for excited states via this updating method, 
we formally introduce the auxiliary $\zeta$-matrix ($\bm{\zeta}$) for a system with M orbitals and N valence electrons ($M > N$), 
which is the transformation matrix from $a_1, a_2, \cdots, a_{N + 1}$ to $a_{N+1}, a_{N+2}, \cdots, a_M$:
\begin{align}
\begin{split}
\begin{bmatrix}
a_{N+1} \\ a_{N+2} \\ \vdots \\ a_M
\end{bmatrix}
=
\begin{bmatrix}
0 & 0 & \cdots & 1 \\
\zeta_{N+2, 1} &  \zeta_{N+2, 2}& \cdots & \zeta_{N+2, N+1} \\
\vdots & \vdots & & \vdots \\
\zeta_{M, 1} &  \zeta_{M, 2}& \cdots & \zeta_{M, N+1}
\end{bmatrix}
\begin{bmatrix}
a_{1} \\ a_{2} \\ \vdots \\ a_{N+1}
\end{bmatrix}
\end{split}
\label{eq:zeta_def}
\end{align}
Rewrite the above matrix multiplication in a compact form, we have $\bm{A}^\text{new} = \bm{\zeta} \bm{A}^\text{ref}$,
where $(\bm{\zeta})_{ij} = \zeta_{N + i, j}$.
Then $\bm{\zeta}$ can be obtained easily via matrix inversion and multiplication
\begin{align}
\begin{split}
\bm{\zeta} = \bm{A}^\text{new} (\bm{A}^\text{ref})^{-1}
\end{split}
\label{eq:zeta_cal}
\end{align}
Note that $\bm{A}^\text{ref}$ is a $(N+1)\times(N+1)$ matrix and we find it is typically invertible in practical calculations.

$\bm{\zeta}$ is of size $(M - N)\times (N + 1)$.
Its column indices map onto the lowest-lying $(N+1)$ orbitals
while the row indices map onto the $(N+1)^{th}$ to $M^{th}$ orbitals.
A $f^{(n)}$ determinant can be obtained from the product of $A^\text{ref}$,
a $n\times n$ minor of $\bm{\zeta}$,
and an overall $\pm$ sign due to permutation of rows (trivial to consider in the single-determinant case).
The last column of this $n\times n$ minor must be taken from the \emph{last} (the $(N+1)^{th}$) column of $\bm{\zeta}$, 
because there are $n$ electron orbitals and $(n - 1)$ hole orbitals in a $f^{(n)}$ configuration,
and the extra one electron can be viewed as removed from the hole on the $(N+1)^{th}$ orbital.
The $n \times n$ minor reflects the interference effect of $n!$ pathways to access the $f^{(n)}$ configuration via permuting $n$ empty orbitals.
The rows (columns) of the $n\times n$ minor indicate the electrons (holes) that are excited in the given $f^{(n)}$ configuration:
the minor formed by rows $i_1, i_2, \cdots, i_n$ ($i_1 < i_2 < \cdots <  i_n$) and columns $j_1, j_2, \cdots, j_{n-1}, N+1$ ($j_1 <  j_2 < \cdots <  j_{n-1} <  N+1$) 
corresponds to the configuration $(c_0, v_1, c_1, \cdots, v_{n - 1}, c_{n - 1}) = (i_1 + N,  j_{n-1},  i_2 + N, \cdots, j_1,  i_n + N)$.
%The overall $\pm$ sign can be omitted because eventually only the determinants' absolute values will be used. 

\subsection{Pruning the configuration space using the breadth-first search algorithm}
\label{sec:bfs}

With this updating technique, we can access the determinants of many configurations without repeatedly carrying out the full determinant calculation for each.
However, the number of configurations still grows exponentially with the order $n$. 
Even the $f^{(3)}$ group grows rapidly as $M^5$, and a system with $M=10^3$ orbitals may have $10^{15}$ $f^{(3)}$ configurations. 
The problem now becomes how to efficiently find all the significant minors of  $\bm{\zeta}$ at all orders.
Enumerating all of these minors will definitely be a hard problem that can not be solved within a polynomial time, 
and the question is whether it is necessary to visit all of them.
In fact, we find that for the systems studied in this work (introduced in Sec. \ref{sec:results}) $\bm{\zeta}$ is sparse, with its non-vanishing elements concentrated in some regions, as will be shown in Sec. \ref{sec:comp}.
A more efficient algorithm should be possible given the sparsity of $\bm{\zeta}$.
 
To make the best use of the sparsity of $\bm{\zeta}$, we investigate its minor determinants in a bottom-up and recursive manner.
According to the Laplace (cofactor) expansion, an $n \times n $ determinant can be expanded into a weighted sum of $n$ minors of size $(n-1)\times(n-1)$.
The $n \times n $ determinant is non-vanishing only when at least one of these $(n-1)\times(n-1)$ minors is non-vanishing.
Physically, this means that a transition to an $f^{(n)}$ configuration is only probable when at least one of its parent $f^{(n-1)}$ configurations is probable, 
otherwise the transition to that $f^{(n)}$ configuration is forbidden.
Assume that $\bm{\zeta}$ is sparse, and one can keep a short list of non-vanishing $(n-1)\times(n-1)$ minors.
When proceeding to $n^{th}$ order, one can construct the $n \times n$ determinant from the short list of non-vanishing $(n-1)\times(n-1)$ minors instead of exhaustively listing all of them. 

%The sparsity is defined as follows. 
%Assume $\zeta_m = ||\bm{\zeta}||_\infty = \max{|\zeta_{ij}|}$. 
%If $|\zeta_{ij}| < r_\text{th}\zeta_m$, where $r_\text{th}$ is a relative threshold for small matrix elements,
%then $\zeta_{ij}$ will be treated as a $0$.
%We can estimate the impact of replacing a small $\zeta_{ij}$ with $0$.
%The largest amplitude of $f^{(n)}$ is of the order of $\zeta^n_m$.
%The contribution from $\zeta_{ij}$ coupled to a $(n-1) \times (n-1)$ minor is at most $|\zeta_{ij}|\zeta_m^{(n-1)}$. 
%The intensity resulting from superposition is $|\zeta^n_m + \zeta_{ij} \zeta^{(n-1)}_m|^2 = |\zeta_m|^{2n} + 2 \text{Re}|\zeta_{ij} \zeta_m^{(2n-1)} |+ \mathcal{O}(|\zeta_{ij}|^2)$,
%where the $\mathcal{O}(|\zeta_{ij}|^2)$ term is $\sim |\zeta_{ij}| |\zeta_m|^{(2n-1)}$, which is $\sim r_\text{th}$ of the maximal intensity $I_m \sim \zeta^{2n}_m$ of $f^{(n)}$.
%{\bfseries [I'm not sure I fully understand what is written here.]}

\begin{figure*}
\centering
  \includegraphics[angle=0, width=0.90\linewidth]{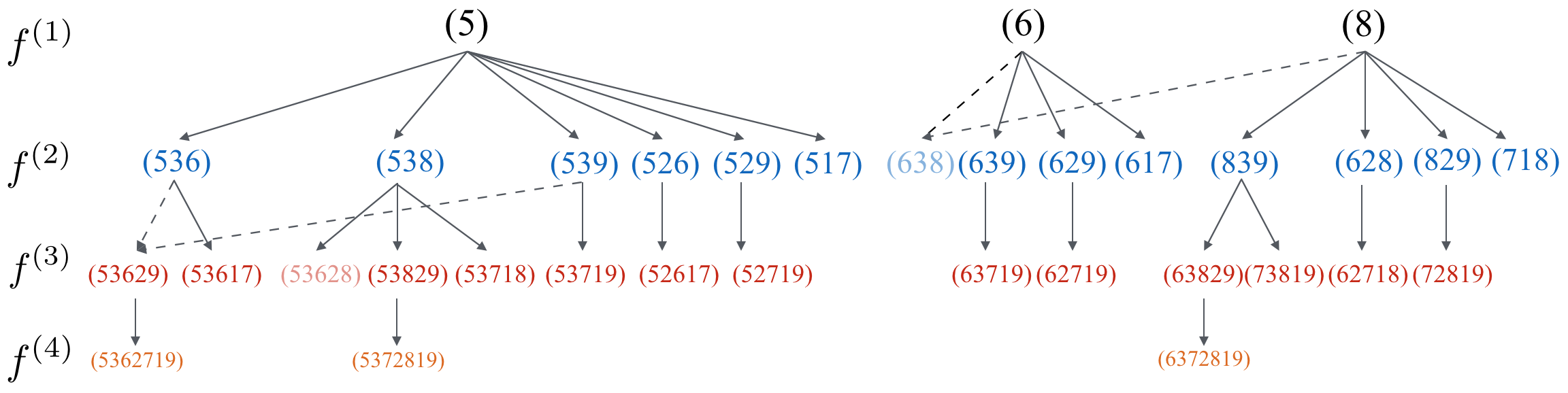}
  \caption{The search tree in the BFS algorithm for finding all nontrivial minors of $\bm{\zeta}$. 
  The digits in the bracket denote the configuration, e.g., $(53629)$ means $(c=5, v_1 = 3, c_1 = 6, v_2 = 2, c_2 = 9)$.
  The semi-transparent configurations are discarded in the search process so that they don't spawn any child configuration.
   }
  \label{fig:tree_bfs}
\end{figure*}

This recursive construction of $n$-minors from the $(n-1)$-minors leads us to an ultimate improvement to the efficiency of the determinantal approach. 
We employ the breadth-first search (BFS) algorithm to enumerate all important minors of $\bm{\zeta}$.
A $f^{(n)}$ configuration can be considered as a descendent of $f^{(n-1)}$ via creating one more \ehpair{} pair with the $f^{(n-1)}$ configuration. 
Through arranging $f^{(n)}$ according to this inheritance relation, a tree-like structure of the many-body expansion is formed, as illustrated in Fig. \ref{fig:tree_bfs}.
The BFS algorithm visits this tree-like structure in ascending order of $f^{(n)}$.
Note that a $f^{(n)}$ configuration can be accessed from its multiple $f^{(n-1)}$ parents via different pathways.
If these pathways to the $f^{(n)}$ configuration interfere destructively such that the transition amplitude is vanishingly small,
the BFS algorithm will discard this $f^{(n)}$ configuration, hence reducing the search space for the next order.
Here is the detailed algorithm

\begin{algorithm}[H]
\caption{Breadth-First Search for Pathways}\label{algo:bfs}
\begin{algorithmic}[1]
\State initialize $f^{(1)}$ * \label{algo:bfs:f1}
\State $n \gets 2$
\Repeat
\For{$f \in f^{(n - 1)}$}
\State extract the indices of $f$: $(c_0, v_1, c_1, \cdots, v_{n - 2}, c_{n - 2})$ 
\ForAll{$\zeta_{cv}$ satisfying $|\zeta_{cv}| > \zeta_\text{th}$} *\label{algo:bfs:zth}
\If{$c\notin \{c_0, c_1, \cdots, c_{n - 2}\} $ and $v < v_{n - 2}$} *\label{algo:bfs:v}
%\State Obtain a new configuration $f' = (c'_0, v'_1, c'_1, \cdots, v'_{i - 1}, c'_{i - 1})$ at the order of $f^{(i)}$
\State Obtain a composite index at $f^{(n)}$ order:
\Statex  \hspace{\algorithmicindent} \hspace{\algorithmicindent} \hspace{\algorithmicindent} $\: f' \gets (c'_0, v'_1, c'_1, \cdots, v'_{n - 1}, c'_{n - 1})$ *\label{algo:bfs:index}
\If{$f' \notin f^{(n)}$}
\State Add $f'$ to $f^{(n)}$
\State $A^{f'} \gets 0$
\State $E_{f'}\gets E_f + (\tilde{\varepsilon_c} - \tilde{\varepsilon_v})$
\EndIf
\State $A^{f'} \gets A^{f'} + (-1)^p \zeta_{cv} A^f $ *\label{algo:bfs:af}
\EndIf
\EndFor
\EndFor
\For{$f \in f^{(n)}$}
\If{$|A^f|^2 < I_\text{th}$} 
Delete $f$ *\label{algo:bfs:ith}
\EndIf
\EndFor
\State Calculate the spectral contribution from $f^{(n)}$
\State $n \gets n + 1$
\Until{the spectrum converges}
\end{algorithmic}
\end{algorithm}

Below are further instructions on the lines marked by asterisks.
\begin{enumerate}
  \item[L\ref{algo:bfs:f1}:]
  $A^f$ of $f^{(1)}$ can be simply taken from the nonzero matrix elements on the last column of $\bm{\zeta}$.
   \item[L\ref{algo:bfs:zth}:]
   $\zeta_\text{th}$ is a threshold for small matrix elements. 
   One can set $\zeta_\text{th} = r_\text{th} \zeta_m$,
   where $\zeta_m\equiv \max{|\zeta_{ij}|} $ and $r_\text{th}$ is a user-defined relative threshold.
  \item[L\ref{algo:bfs:v}:]
  The $n\times n$ determinant of $f^{(n)}$ is constructed via a Laplace expansion along its \emph{first column}.
  $v < v_{n - 2}$ ensures the chosen matrix element $\zeta_{cv}$ is always on the first column of the $n\times n$ determinant.
  \item[L\ref{algo:bfs:index}:]
  Compare to $f$, $f'$ contains one more \ehpair{} pair labeled by $c$ and $v$.
  Because we require the ordering of $c_0 < c_1 < \cdots $ and $v_1 > v_2 > \cdots$ for unique indexing,
  the new index $(c'_0, v'_1, c'_1, \cdots, v'_{n - 1}, c'_{n - 1})$ must obey the same order.
  The new sequence $(c'_i)$ can be obtained by this procedure: 
  first place the new $c$ in front of the $(c_i)$ sequence of $f^{(n-1)}$ that is already increasingly sorted,
  and then shift $c$ to the right by swapping indices till the whole sequence is also sorted.
  Define $p$ to be the number of swaps performed for deciding signs.
  $(v'_i)$ can be obtained simply by placing $v$ at the end of $(v_i)$.
  \item[L\ref{algo:bfs:af}:]
  $(-1)^p \zeta_{cv}$ is the cofactor of the Laplace expansion of a $f^{(n)}$ determinant.
  where $p$ is the proper position for inserting $c$ into $(c_i)$, as defined above.
  At the end of the $\zeta_{cv}$ loop, there are at most $n$ contributions to the total amplitude $A^{f'}$ of a specific $f^{(n)}$ configuration, 
  corresponding to the transition amplitudes of $n$ different pathways from its parent $f^{(n-1)}$ configuration.
  \item[L\ref{algo:bfs:ith}:]
  $I_\text{th}$ is a threshold for removing state with small oscillator strengths.
  Similar to $\zeta_\text{th}$, $I_\text{th}$ can be set to $I_\text{th} = R_\text{th} I_m$, 
  where $I_m$ is the maximal oscillator strength and $R_\text{th}$ is a user-defined relative threshold.
  $I_m$ can be chosen to be the maximal intensity within the $f^{(1)}$ group which typically have the strongest oscillator strengths among all $f^{(n)}$ groups. 
  $R_\text{th}$ can be related to the previously defined relative matrix-element threshold $r_\text{th}$.
  If the contribution from a small $a_{cv}$ were not added to $A^{f'}$, 
  its intensity would be $|A^{f'} - (-1)^p \zeta_{cv} A^f |^2 = | A^{f'} |^2 - 2 (-1)^p \text{Re}[\zeta_{cv} A^{f'} (A^f)^*] + \mathcal{O}(|\zeta_{cv}|^2)$.
  Replacing $\zeta_{cv}$ with $0$ will lead to an error of $\sim |\zeta_{cv} | |A^{f'}| |A^f | \leq r_\text{th} |\zeta_m | I_m$.
  Therefore, choosing a $r_\text{th}$ such that $R_\text{th} \sim \zeta_m  r_\text{th}$ can guarantee error in intensities smaller than $I_\text{th} = R_\text{th} I_m$.
  In practice, one can lower $R_\text{th}$ till convergence is achieved.
  
\end{enumerate}

The detailed implementation of this search algorithm can be found at Ref. \cite{mbxaspy} within an open-source PYTHON simulation package.

Here we demonstrate the BFS algorithm with a toy model with $M = 9$ orbitals and $N = 4$ valence electrons. 
%We have in hand the reference determinant, $A^\text{ref}=\det A^\text{ref}$.
Suppose $\bm{\zeta}$ of the system is
\begin{align}
\begin{split}
\begin{bmatrix} %{c(ccccc)}
   & 1 & 2 & 3 & 4 & 5 \\
5 & 0 & 0 & 0 & 0 & 1 \\
6 & 0 & \zeta_{62} & \zeta_{63} & 0 & \zeta_{65} \\
7 & \zeta_{71} & 0 & 0 & 0 & 0 \\
8 & 0 & 0 & \zeta_{83} & 0 & \zeta_{85} \\
9 & 0 & \zeta_{92} & \zeta_{93} & 0 & 0 \\
\end{bmatrix}
%\kbordermatrix{
%   & 1 & 2 & 3 & 4 & 5 \\
%5 & 0 & 0 & 0 & 0 & 1 \\
%6 & 0 & \zeta_{62} & \zeta_{63} & 0 & \zeta_{65} \\
%7 & \zeta_{71} & 0 & 0 & 0 & 0 \\
%8 & 0 & 0 & \zeta_{83} & 0 & \zeta_{85} \\
%9 & 0 & \zeta_{92} & \zeta_{93} & 0 & 1 \\
%}
\end{split}
\label{eq:zeta_demo}
\end{align}
The BFS algorithm for this example of $\bm{\zeta}$ is carried out as follows:

First, the non-zero $f^{(1)}$ configurations are initialized, (5), (6), and (8), 
whose determinants are simply the matrix elements: $1$, $\zeta_{65}$, and $\zeta_{85}$, respectively. 
These configurations are considered as the roots of the BFS trees,
as is shown in Fig. \ref{fig:tree_bfs}.

Next, the $f^{(2)}$ configurations are constructed based on the obtained $f^{(1)}$ configurations.
Take the configuration $(8)$ in $f^{(1)}$ for example.
There are 5 non-zero matrix elements that are to the left of $(8)$ and are not on the same row as $(8)$,
which are $\zeta_{63}$, $\zeta_{93}$, $\zeta_{62}$, $\zeta_{92}$, and $\zeta_{71}$.
Paired up with these matrix elements, the $(8)$ configuration spawns 5 $f^{(2)}$ configurations:
$(638)$, $(839)$, $(628)$ , $(829)$, and $(718)$ (comma omitted due to the single-digit indices).
Likewise, $(5)$ and $(6)$ spawn 6 and 4 $f^{(2)}$ configurations respectively.

Both $(6)$ and $(8)$ give rise to $(638)$ and the contributions from the $f^{(1)}$ configurations are merged:
A(638) = $\zeta_{63}\zeta_{85} - \zeta_{83} \zeta_{65}$.
The two possible pathways are: 
(1) the core electron is first promoted to orbital $6$ and then coupled with the \ehpair{} pair formed by orbital $3$ and $8$;
(2) the core electron is first promoted to orbital $8$ and then coupled with the \ehpair{} pair formed by orbital $3$ and $6$.
If $A(638)$ is vanishingly small ($\zeta_{63}\zeta_{85}$ happens to be close with $\zeta_{83} \zeta_{65}$) due to the destructive interference of the two pathways, 
Then $(638)$ will be removed from the $f^{(2)}$ list because it cannot contribute to the transition amplitude of any higher-order configuration. 
When the search process for $f^{(2)}$ is completed, $13$ nontrivial configurations are found.

Proceeding to the third order, the 13 $f^{(2)}$ configurations spawn $14$ $f^{(3)}$ configurations.
Paring $(536)$ with $\zeta_{92}$ and $(539)$ with $\zeta_{62}$ both lead to $(53629)$, whose determinant is $\zeta_{62}\zeta_{93} - \zeta_{63}\zeta_{92}$.
Paring $(538)$($=\zeta_{83}$) with $\zeta_{62}$ leads to $(53628)$ (the other two pathways are forbidden because $\zeta_{52} = \zeta_{82} = 0$).
If $\zeta_{83}$ and $\zeta_{62}$ are small numbers such that their product is smaller than the specified threshold $I_\text{th}$, 
then $(53628)$ will be removed from $f^{(3)}$.
The above process can be repeated until all new determinants are small enough or no new determinants can be found.
 
If one brute-forcely enumerates all possible determinants, 
there are $\binom{5}{2}\binom{5}{1} = 50$ $f^{(2)}$ and $\binom{5}{3}\binom{5}{2} = 100$ $f^{(3)}$ determinants to examine for the above $\zeta$-matrix.
By contrast, the BFS algorithm only visits the nontrivial determinants and only $14$ $f^{(2)}$ and $14$ $f^{(3)}$ determinants are computed.

\subsection{Incorporating first-principles calculations into the determinant formalism}
\label{sec:dscf}
In the above sections, we have demonstrated an efficient solution to the MND model using many-body wave functions
for simulating x-ray transition amplitudes.
However, in order to simulate reliable x-ray spectra without fitting parameters from experimetns, 
we still need accurate approximations to the initial and final states and their energies.
To this end, we rely on DFT calculations to obtain the KS eigenstate energies (for $\tilde{\varepsilon}_f$) and wavefunctions (for both $|\tilde{\psi}_i\rangle$ and $|\psi_j\rangle$) 
as input for constructing the transformation matrix $A^f_c$ (Eq. (\ref{eq:afc}))
and computing the energies of many-electron excited states (Eq. (\ref{eq:Ef})). 

For the final state, we employ the standard $\Delta$SCF core-hole approach to obtain the KS orbitals and eigenenergies.
The core-excited atom is treated as an isolated impurity embedded in the pristine system, 
and typical supercell settings for finite \cite{england2011hydration} and extended \cite{drisdell2013probing, velasco2014structure, pascal2014x, liang2017accurate} systems can be employed.
To simulate a electron-core-hole pair, 
the core-excited atom is modeled by a modified pseudopotential with a core hole,
and an electron is added to the supercell system and constrained to one specific empty orbital.
In principle, a $\Delta$SCF iteration needs to be performed for each case of constraint occupancy (for all $f=(1,2,\dots,N,c)$)
which may lead to an expensive computational cost.
As a trade-off, the electron is only placed onto the lowest unoccupied orbital ($f=(1,2,\dots,N,N+1)$), 
which we have dubbed the excited-state core-hole (XCH) method.
After the $\Delta$SCF calculation is done, the KS equation with a converged charge density is used for $\mathcal{H}_f$. 

Another important variation of the XCH method is the full core-hole (FCH) approach, 
in which the ground-state occupation $f=(1,2,\dots,N)$ without the additional electron is used.
The advantage of FCH is that it does not bias towards the lowest excited state and treat all of them on equal footing.

For the initial state, the same supercell as in the final state is used except that the core-excited atom is replaced by a ground-state atom,
using the occupation $f=(1,2,\dots,N)$.
A standard DFT calculation can be done to obtain the KS orbitals $|\psi_i\rangle$.

With the KS orbitals $|\tilde{\psi}_i\rangle$ and $|\psi_j\rangle$ obtained from the $\Delta$SCF core-hole calculation,
we can compute the orbital overlap integral $\xi_{ij} = \langle \psi_j |\tilde{\psi}_i \rangle$ for computing the determinantal amplitudes.
To reduce computational cost, %for systems involving $d$ orbitals, 
we employ projector-augmented-wave (PAW) form of ultrasoft pseudopotentials \cite{kresse1999ultrasoft, blochl1994projector, taillefumier2002x} to model electron-ion interactions.
The excited atom potential has deeper energy levels and more contracted orbitals so its PAW construction differ from the ground-state atom.
In Appendix \ref{sec:paw}, we derive a formalism to calculate expectation values between 
$|\psi_j\rangle$ of the ground-state system and
$|\tilde{\psi}_i\rangle$ of the core-excited system.
Initial dipole matrix elements $\langle \psi_c | o | \psi_h \rangle$ are also evaluated within this PAW formalism \cite{blochl1994projector, taillefumier2002x}.

As in typical DFT impurity calculations, some low-lying excited states in the core-hole approach could be bound to the core-excited atom, resembling mid-gap localized electronic states near an impurity.
In this situation, the electronic structure is well described by using a single k-point (the $\Gamma$ point) to sample the Brillouin zone (BZ).
However, for the purpose of spectral simulations, which include delocalized scattering states well above the band edges, 
we find that employing $k$-point sampling is necessary to improve the accuracy of the calculated line shape.
Therefore, we perform the determinantal calculation individually for each k-point and 
take the k-point-weighted average spectrum as the final spectrum.
The band structure and orbitals are interpolated accurately and efficiently using an optimal basis set proposed by Shirley \cite{shirley1996optimal, prendergast2009bloch},
whose size is much smaller than a plane-wave basis. 
In the Shirley construction, the periodic parts of Bloch wave functions across the first BZ are represented using a common basis which spans the entire band structure.
Because there is only one optimal basis to represent the Bloch states for all k-points, 
the overlap $\xi$-matrix for every k-point can be computed quickly as in Appendix \ref{sec:obf}.

After the XAS is calculated by the first-principles determinantal approach,  
the established formation-energy calculation can be adopted to align spectra for core-excited atoms in different chemical contexts, using the XCH method to determine the excitation energy of the first transition \cite{england2011hydration, jiang2013experimental}.

Although there is no valence \eepair{} interaction terms in the MND theory, 
which results in a single-determinant solution to the many-body wave functions,
we argue that this first-principles determinantal approach \emph{does not} entirely neglect valence \eepair{} interactions.
The self-consistent-field (SCF) procedure in the DFT updates the total charge density and KS orbitals simultaneously, 
and hence takes into account some degree of valence electronic screening. 
That said, the $\Delta$SCF approach should lead to a more realistic equilibrium total charge density 
for both the ground state and x-ray excited states,
whereas the charge density in MBPT is only treated perturbatively.
Moreover, a more accurate charge density may lead to a better approximation to quasiparticle (QP) wave functions.
In fact, KS orbitals based on a converged SCF are often employed in MBPT to construct the Green's functions and compute optical oscillator strengths \cite{hybertsen1986electron, rohlfing2000electron, onida2002electronic},
which is typically a good approximation within the Fermi-liquid picture.
Finally, corrective DFT (DFT + $U$ or DFT with exact-exchange functionals) or the self-consistent $GW$ approximation \cite{bruneval2006effect, van2006quasiparticle, kang2010enhanced, sun2017x}
can also improve QP energies and wave functions to be used in the determinantal approach.
We imagine the many-body effects captured in this framework can be described by the bolded version of the MND diagrams in Fig. \ref{fig:lcc_2nd}
in which all the Green's functions become dressed, 
and the bare Coulomb lines are replaced by the screened core-hole potential described with the chosen exchange-correlation functional.

\subsection{Comparison with the one-body $\Delta$SCF core-hole approach}
\label{sec:onebody}

Before the determinantal formalism, the many-body transition amplitudes in the $\Delta$SCF core-hole approach are often approximated with 1p matrix elements
\begin{equation}
\langle\Psi_f| \bm{\epsilon}{\cdot}\bm{R} |\Psi_i\rangle \approx  S\langle\tilde{\psi_f}| \bm{\epsilon}{\cdot}\bm{r} |\psi_h\rangle
\label{eq:dscf_mat_elem}
\end{equation}
where the core orbital $|\psi_h\rangle$ is in the initial state while the electron orbital $|\tilde{\psi}_f\rangle$ is in the final states, 
both of which can be taken from DFT calculations.
$S$ represents the response of rest of the many-electron system (excluding the electron-core-hole pair) due to the core hole,
and it is normally assumed to be a constant for ease of calculations.
This 1p form of the matrix element implies that:
(a) the transition from the initial core level $|\psi_h\rangle$ to the final electron orbital $|\tilde{\psi}_f\rangle$ occurs instantaneously 
with the response of many other electrons in the system, with no particular time ordering;
(b) the core-level transition and the many-electron response are not entangled.
This is also the so called sudden or frozen approximation. 

We know that from the diagrammatic interpretation of the x-ray many-body processes in Fig. \ref{fig:lcc_2nd}, 
the photon first decays into an initial-state \ehpair{} pair instantaneously, 
and then the other electrons see the core-hole potential and begin to relax over a finite period of time.
This physical reality can also be seen in the determinant formalism, in which
the core hole is only coupled to an initial-state orbital, and the subsequent many-electron response is described by the determinantal amplitude.
So the question is why the simpler 1p matrix element in the frozen approximation still works for a good number of systems in the past.

In this section, we approach this question theoretically by relating the determinantal amplitude to the 1p matrix element.
To do this, we first express the $(N+1)\times(N+1)$ determinantal amplitude $A^{f}_c$ in terms of is $N\times N$ minors 
(wavefunction overlaps of $N$-electron systems, such as $S$) by Laplace expansion along its last column
\begin{align}
\begin{split}
A^{f}_c = \sum^{N+1}_{i=1} M^f_{i}  \xi_{f_i, c} 
\end{split}
\label{eq:afc_minor}
\end{align}
where $\xi_{f_i, c}$ are the matrix elements on the last column of $A^f_c$ as in Eq. (\ref{eq:afc}) and $M^f_{i}$ is the minor complementary to $\xi_{f_i, c}$.
Since in the one-body core-hole approach only the $f^{(1)}$ terms are summed, we limit our analysis here to the many-body $f^{(1)}$ terms and condense the configuration tuple into a single index:
$(1, 2, \cdots, N, f) \mapsto f$.
Then the matrix elements $\langle \Psi_f | O | \Psi_i \rangle$ can be written as
\begin{align}
\begin{split}
&\sum_{c \in \text{empty}} (A^{f}_c)^* \langle \psi_c | o | \psi_h \rangle\\
=& (M^f_{N+1})^* \sum_{c \in \text{empty}} \langle \tilde{\psi}_f  | \psi_c \rangle \langle \psi_c | o | \psi_h \rangle\\
+ & \sum^{N}_{i=1}(M^f_{i})^* \sum_{c \in \text{empty}}   \langle \tilde{\psi}_i  | \psi_c \rangle \langle \psi_c | o | \psi_h \rangle 
\end{split}
\label{eq:mat_decomposed}
\end{align}

First, for systems with significant band gaps (insulators and semiconductors), we could expect that the overlap of the occupied final state orbitals with the unoccupied initial state orbitals could be quite small. For many orbitals unaffected by the localized core-hole perturbation, for example, we might expect the final state occupied orbitals to closely resemble their initial state counterparts, which would render $\langle \tilde{\psi}_v  | \psi_c \rangle$ identically zero by orthogonality. Therefore, the sum over $v$ in Eq. (\ref{eq:mat_decomposed}) may only be significant in cases where the transformation matrix $\mathbf{\xi}$ indicates mixing of unoccupied initial state character into the occupied final state orbitals, which might easily be the case for orbitals close to the Fermi level in a metal or otherwise open-shell system.

The first term in Eq. (\ref{eq:mat_decomposed}) is more directly relevant to our previous one-body approximation. Here, $M^f_{N+1}$ is the minor of $(\xi_{ij})_{N\times N}$, the transformation matrix without its $(N+1)^\textrm{th}$ column and row. It reflects the $N$-electron many-body overlap between the initial and final state occupied orbitals and should reflect the extent to which the electron density is modified by the core-hole perturbation.
Since $M^f_{N+1}$ does not depend on $f$,
we can relate it to the many-body prefactor that appears in the final-state rule of Eq. (\ref{eq:dscf_mat_elem}): $S=(M^f_{N+1})^*$.
Using the completeness relation: $\sum_{c \in \text{empty}}  | \psi_c \rangle \langle \psi_c | =  \mathbb{1} - \sum_{v \in \text{occ}}  | \psi_v \rangle \langle \psi_v |$, the first term in the expansion of Eq. (\ref{eq:mat_decomposed}) can be expressed as
\begin{align}
\begin{split}
S \big[ \langle \tilde{\psi}_f | o | \psi_h \rangle -  \sum_{v \in \text{occ}} \langle \tilde{\psi}_f  | \psi_v \rangle \langle \psi_v | o | \psi_h \rangle \big]
\end{split}
\label{eq:compare_to_f}
\end{align}
If it happened that $\langle \psi_v | \tilde{\psi}_f \rangle = 0$, then this expression would amount to the final state matrix element as defined in the one-body final-state rule (Eq. (\ref{eq:dscf_mat_elem})). By the same arguments made above, for systems with limited mixing of orbital character across a significant band gap, then we might easily expect orthogonality (zero overlap) between occupied initial state and unoccupied final state orbitals. By the same token, we should be wary of limitations in the one-body approach when this is not the case.

It appears useful to focus on $\langle \psi_v | \tilde{\psi}_f \rangle$ to reveal the role of  hybridization in modulating near-edge spectral intensity.
To quantify the contribution of the second term in Eq. (\ref{eq:compare_to_f}), we introduce the projection spectrum 
\begin{align}
\begin{split}
\sigma_{fi} (E) 
&= \sum_{f} | \langle \tilde{\psi}_f | P_c o|\psi_h\rangle|^2 \delta(E - \tilde{\varepsilon}_f) \\
\end{split}
\label{eq:sigma_fi}
\end{align}
in which the single index $f$ sums over all empty final-state orbitals, and $P_c \equiv \sum_{c \in \text{empty}} | \psi_c \rangle \langle \psi_c |$.
The matrix element is nothing but Eq. (\ref{eq:compare_to_f}) or the first term in Eq. (\ref{eq:mat_decomposed}) with $S=1$.
However, it is easier to calculate Eq. (\ref{eq:compare_to_f}) because summation over all empty orbitals is avoided. 

\section{RESULTS AND DISCUSSION}
\label{SEC:results}
\subsection{Applications to transition-metal oxides}
\label{sec:results}
In this section, we discuss an important application of the determinantal approach to computing core-excited state transition amplitudes, that is, to predict the x-ray absorption spectra (XAS) for transition metal oxides (TMOs).
This is also our original motivation for proposing the determinantal approach \cite{liang2017accurate}, which can be used to overcome the deficiency of the one-body core-hole approach.
It has been found for a number of TMOs, that the one-body approach systematically underestimates the intensity of near-edge features at the O $K$ edge that correspond to orbitals with hybridization between oxygen $p$-character and TM $3d$-character.
This underestimation can prevent reliable interpretations of the X-ray absorption spectra for this important class of materials.

We use the newly developed determinantal approach to predict the XAS for eight TMOs: 
the rutile phase of 
\ce{TiO2},  
\ce{VO2} ($>340$ K), 
and \ce{CrO2}, 
the corundum \ce{Fe2O3},
the perovskite \ce{SrTiO3},
\ce{NiO}, 
and \ce{CuO}.
\ce{SiO2} is also chosen for a comparative study.
Their experimental XAS are extracted from Refs. \cite{yan2009oxygen, koethe2006transfer, stagarescu2000orbital, shen2014surface, zhu2012bonding, lin2014hierarchically, jiang2013experimental, dudarev1998electron, ma1992soft}.
The chosen TMOs cover a wide range of electronic and magnetic properties and therefore they are used as benchmark materials for the determinantal approach.
%How severe the one-body approach would fail in predicting the near-edge lineshape is closely tied to the band gap and the nature of the bonding of the system, as we will show next.

The \ce{O} $K$ edges are investigated here, i.e., the transitions from the \ce{O} $1s$ level to $np$ shells.
For TMOs, the \ce{O} $2p$ orbitals are covalently hybridized with the transition metal $3d$ orbitals, 
and hence the \ce{O} $K$-edge spectra can serve as an informative and sensitive probe for the $d$-electron physics \cite{de1989oxygen, yabuuchi2011detailed, hu2013origin, suntivich2011design, lin2016metal, luo2016charge, strasser2010lattice, matsukawa2014enhancing, lebens2016direct, de2016mapping}.
Moreover, unlike transition metal $L_{2,3}$ edges ($2p$-to-$3d$ transitions), in which atomic multiplet effects split spectral features into many closely space lines\cite{de2008core}, the \ce{O} $K$ edges can provide a picture of the electronic density-of-states related to the $d$ shell more easily interpretable in terms of band theory or effective 1p states. 

\begin{figure*}
\centering
\sidesubfloat[(a)]{
  \includegraphics[angle=0, width=0.9\linewidth]{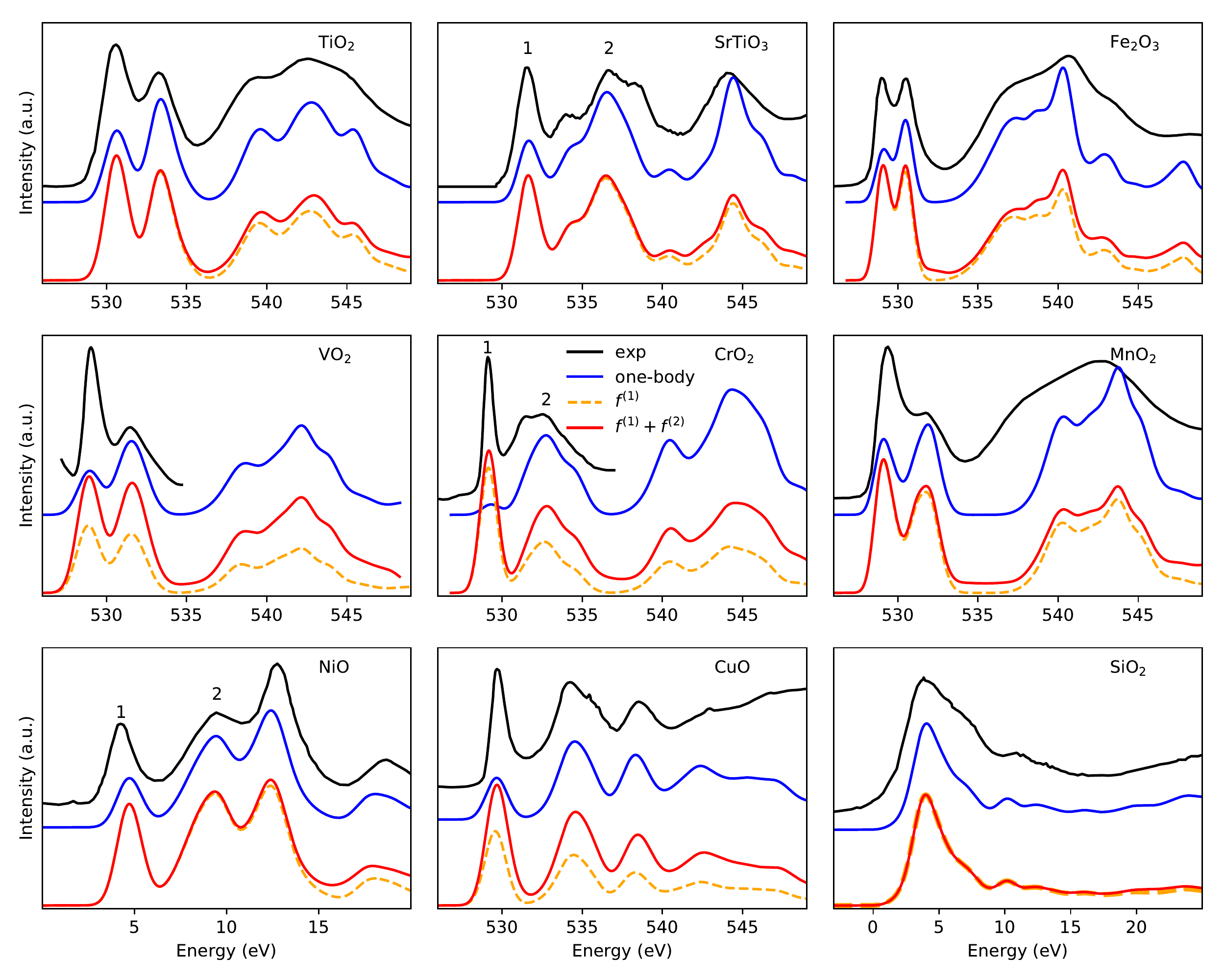}
  }\\
\sidesubfloat[(b)]{
  \includegraphics[angle=0, width=0.3\linewidth]{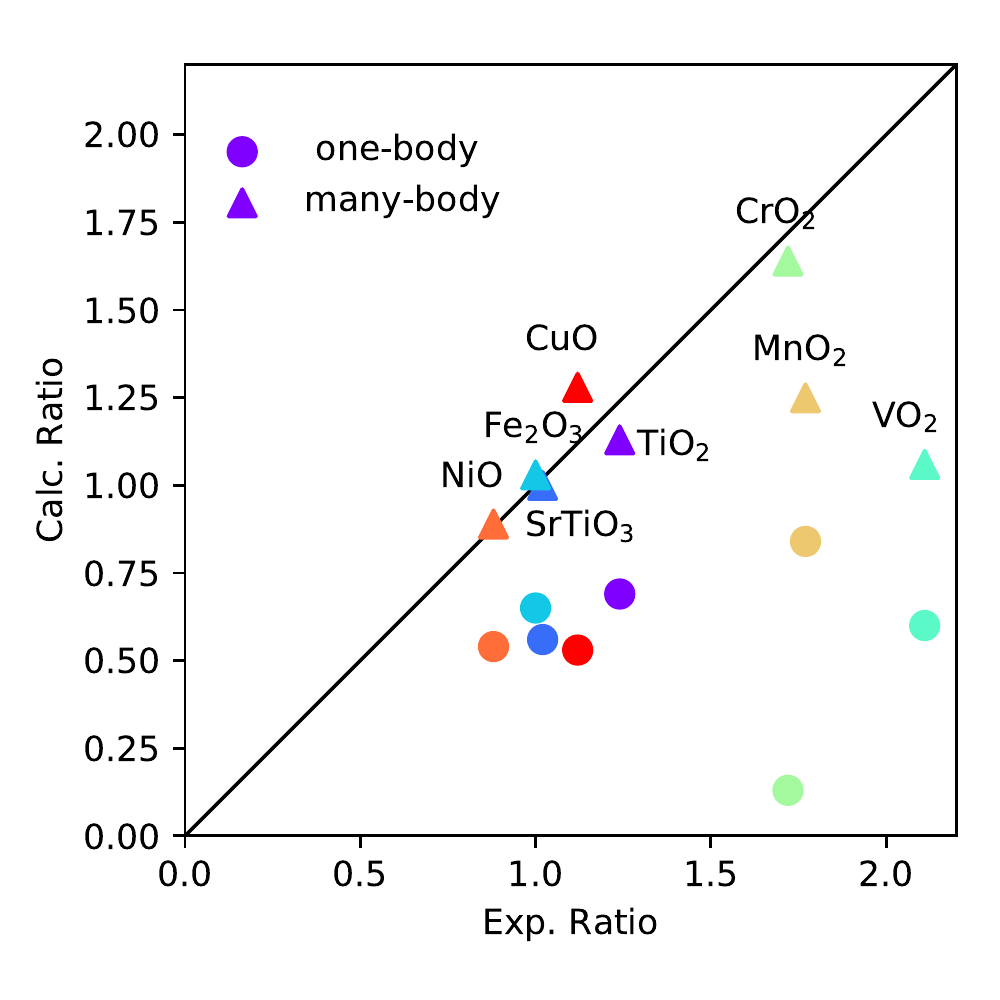}
 }
 \sidesubfloat[(c)]{
  \includegraphics[angle=0, width=0.5\linewidth]{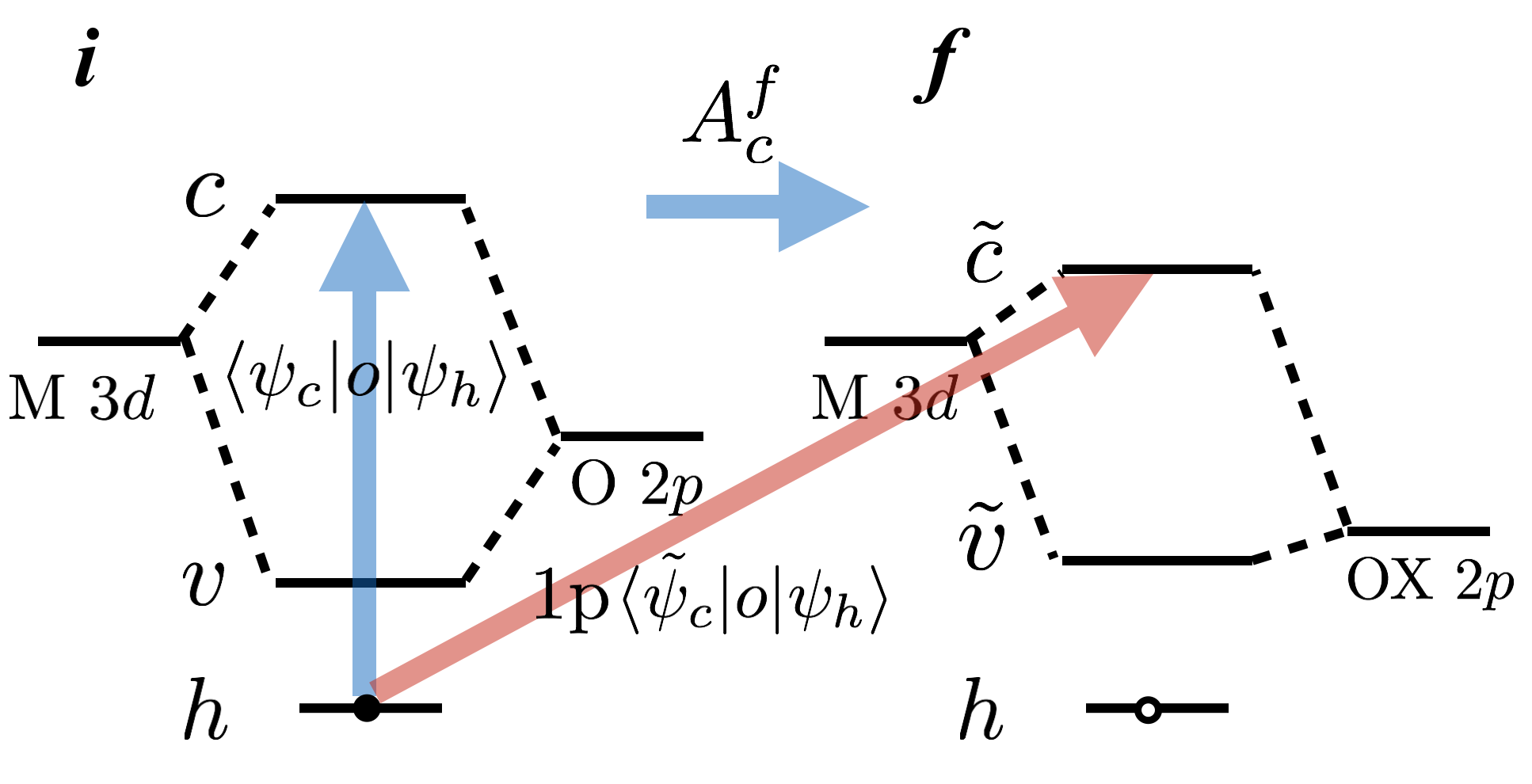}
 }
\caption{ (Color online)
\textbf{(a)} XAS for the selected crystal structures obtained from experiments (black),  one-body FCH approach (blue), and the many-electron determinant approach (red) introduced in this work.
The XAS calculated with the $f^{(1)}$ configurations are shown by dashed orange curves.
The energy axes for \ce{NiO} and \ce{SiO_2} are relative.
\textbf{(b)} Comparison of experimental peak intensity ratios compared with the ones predicted by the one-body (circles) and the many-electron (triangles) formalism.
Each color represents the result for one system.
The peak intensity ratio refers to the ratio of the lowest-energy maxima to the second of the spectrum, unless otherwise specified by the numbers in (a). 
The spectra are broadened to the best as compared with experimental broadenings.
\textbf{(c)} Schematics showing how the one-body and the many-electron formalism treats x-ray excitations, using a metal-$3d$-\ce{O}-$2p$ molecular model in both 
the initial (i) and final (f) state. 
The one-body approach mainly relies on the single-particle (1p) matrix element and has skipped (red arrow) the dynamics of the many-electron charge relaxation,
while the many-electron formalism considers the actual multiple-step (blue arrows) excitation process that involves all the electrons in the system.
}
\label{fig:results}
\end{figure*}

The angularly-averaged (except in \ce{CrO2}, where the polarization is perpendicular to the hard axis) \ce{O} $K$-edge spectra for the chosen TMOs are shown in Fig. \ref{fig:results} (a).
The very near-edge part of the spectra, i.e., the spectral features below $535$ eV contain the most useful information for $3d$ material characterization.
For these TMOs, the near-edge spectral fine structure exhibits two main peaks corresponding to the splitting of the $d$-orbitals into a $t_{2g}$ and an $e_g$ manifold in the (quasi-)octahedral crystal field.
Our goal is to produce reliably all the spectral features, especially the very near-edge part, so that one can interpret the spectra on a first-principles basis.
More specifically, we use the ratio of the intensity of the first (lowest-lying) peak to that of the second (unless otherwise specified) as a metric for the accuracy of different levels of approximation.

\floatsetup[figure]{subcapbesideposition=top}

We first calculate the XAS for the chosen compounds using the conventional 1p FCH approach \cite{taillefumier2002x, prendergast2006x, liang2017accurate} described above.
A modified pseudopotential generated with the configuration $1s^1 2s^2 2p^4$ is used for the $1s$-core-excited \ce{O}.
We choose supercell dimensions of approximately $10$\AA{} that is sufficient to separate the effect of the core-hole impurity from its neighboring periodic images. 
The FCH calculations are performed using the DFT+$U$ theory \cite{dudarev1998electron} with the $U$ value adopted from Ref. \cite{wang2006oxidation}.
A uniform $5\times 5 \times 5$ $\bm{k}$-point grid of the supercell BZ is employed to sample a continuous density-of-states at higher energies.
As we have demonstrated by calculations before \cite{liang2017accurate}, the 1p FCH approach universally underestimates the peak intensity ratio for all selected TMOs (blue curves in Fig. \ref{fig:results} (a)).  
This includes the newly added cases: \ce{MnO2}, \ce{NiO}, and \ce{CuO}, where the peak intensity ratios are just $50\%$ of the experimental ones.

The failure of the 1p FCH approach motivated us to use the determinant formalism in Eq. (\ref{eq:afc}) as a better approximation to the dipole matrix elements \cite{liang2017accurate}.
In this work, we implement the determinant approach with the efficient procedures discussed in Sec. \ref{sec:dscf} and the BFS algorithm.
We use exactly the same final-state SCF as in the 1p FCH approach and an initial-state supercell of the same dimensions.
Besides employing the BFS algorithm to reduce the computational cost, we separate the two spin channels to speed up the calculations. 
In the absence of spin-orbital coupling, the $\xi$-matrix is block-diagonalized and each transition either occurs within the spin-up or spin-down manifold, 
and the BFS algorithm can be performed over each spin manifold with a reduced $\xi$-matrix. 
The total absorption spectra can be obtained from combining individual spectra from the two spin channels (for the collinear case) using the spectral convolution theorem in Appendix \ref{sec:conv} 
\begin{align}
\begin{split}
\sigma_{\text{XAS}}(E) &=\int d E' \sigma_{\text{XAS}, \uparrow}(E - E') \sigma_{\text{XPS}, \downarrow} (E') + \{\uparrow \rightleftharpoons\downarrow\}
\end{split}
\label{eq:xas_conv}
\end{align}
where $\sigma_{\text{XAS}, \mu}$ is the XAS of an individual spin channel $\mu$. 
$\sigma_{\text{XPS}, \mu}$ is the core-hole spectral function of spin $\mu$
\begin{align}
\begin{split}
\sigma_{\text{XPS}, \mu} (E)= \sum_{f}\langle \Psi^{N_\mu}_{\mu, f} | \Psi^{N_\mu}_{\mu, i} \rangle \delta(E-(E_{\mu, f} - \min E_{\mu, f}))
\end{split}
\label{eq:ch_A}
\end{align}
Here $|\Psi^{N_\mu}_{\mu, f}\rangle$ ($|\Psi^{N_\mu}_{\mu, i}\rangle$) is the $N_\mu$-electrons many-body wave function
of final state $f$ (initial state $i$) within the spin manifold $\mu$, 
and $N_{\mu}$ is the number of electrons in its initial state ($N_{\uparrow}\neq N_{\downarrow}$ for a ferromagnetic system). 
Because the core-hole spectral function 
is analogous to the corresponding x-ray photoemission spectrum (XPS) \cite{kas2015real}, we dub the former as $\sigma_{\text{XPS}}$ hereafter.
The calculation of $\sigma_{\text{XPS}, \mu}$ entirely resembles that of $\sigma_{\text{XAS}, \mu}$ and the prominent matrix elements $\langle \Psi^{N_\mu}_{\mu, f} | \Psi^{N_\mu}_{\mu, i} \rangle$ are also found by the BFS algorithm as in Sec. \ref{sec:sol}.

The spectra calculated with the determinantal approach up to $f^{(2)}$ order are shown in Fig. \ref{fig:results} (a).
There is substantial improvement in the peak intensity ratios and the overall line shapes for the TMOs being investigated.
In particular, the peak intensity ratios of \ce{TiO2}, \ce{CrO2}, \ce{Fe2O3}, \ce{CuO}, \ce{NiO}, and \ce{SrTiO3} are in excellent agreement with experiments [Figs. \ref{fig:results} (a) and (b) ].
The peak intensity ratio of \ce{VO2} is still underestimated, however, this may be related to missing contributions to the leading edge from the nearby \ce{V} $L$ edge, which is not included in our simulation.\cite{koethe2006transfer}. 
The prediction of the peak intensity ratio of \ce{MnO2} is less satisfactory partly because we simulate its spectrum using a rutile unit cell with colinear antiferromagnetic order, whereas its actual magnetic order is found to be helical and has a larger periodicity \cite{tompsett2012importance, lim2016improved}.
The lack of anisotropy in the Hubbard $U$ interactions in our current calculation may also explain why the simulated spectrum deviates from experiments.
More advanced treatment of strongly correlated materials, using hybrid functionals, for example,\cite{heyd2003hybrid, chai2008long, paier2009cu, paier2008dielectric}, could be coupled with the determinantal formalism to produce more accurate results. 
In principle, any effective 1p orbital basis can be used in this formalism.

\subsection{Origins of XAS intensity underestimation using one-body approaches}
\label{sec:intunder}

In a nutshell, the underestimation of the peak intensity ratios by the one-body approach can be understood from a three energy-level model.
Consider a molecule with one single metal level (M) hybridized with an \ce{O} $2p$ level, plus one \ce{O} $1s$ core level, as is shown in the schematics in Fig. \ref{fig:results} (c). Hybridization within the empty $(c)$ and filled $(v)$ states can be expressed using a unitary transformation of the corresponding atomic orbitals:
$(|\tilde{\psi}_c\rangle, |\tilde{\psi}_v\rangle)^T = R(\theta_i) (|\text{M}_{3d}\rangle, |\text{O}_{2p}\rangle)^T $, where
$R(\theta_i)$ is a 2D rotation matrix
\begin{align}
\begin{split}
R(\theta)=
\begin{bmatrix}
\cos\theta & -\sin\theta \\
\sin\theta & \cos\theta
\end{bmatrix}
\end{split}
\label{eq:model_1}
\end{align}
Initially the system is half filled and its hybridization represented by an angle $\theta_i\in [0,\pi/2]$.
The final state can be expressed likewise using its own angle $\theta_f$: 
$(|\tilde{\psi}_c\rangle, |\tilde{\psi}_v\rangle)^T = R(\theta_f) (|\text{M}_{3d}\rangle, |\text{O}_{2p}\rangle)^T $.
Phenomenologically, we expect the initial and final states to differ in their degree of hybridization of these two atomic levels. 
The core-hole potential lowers the energy of the oxygen $2p$ orbital in the final state, 
enhancing the $ |\text{O}_{2p}\rangle$ component of the occupied final-state orbital $v$ and reducing the same for the unoccupied final-state orbital $c$.
Hence, $0 < \theta_f < \theta_i$.

Within this minimal model of just two electrons, there is only one available core-excited transition, 
i.e., the excitation from $i=(h,v)$ to the final state $f=(\tilde{v},\tilde{c})$.
The exact spectral intensity calculated by the many-electron formalism as in Eq. (\ref{eq:afc}) is
\begin{align}
\begin{split}
|\langle\Psi_f| \bm{\epsilon}{\cdot}\bm{R} |\Psi_i\rangle |^2
&= |\det [ R(\theta_i-\theta_f) ] \langle \psi_c|\bm{\epsilon}\cdot\bm{r}|\text{O}_{1s}\rangle|^2\\
&= |1\times\langle \psi_c|\bm{\epsilon}\cdot\bm{r}|\text{O}_{1s}\rangle|^2\\
&= \sin^2 \theta_i |\langle \text{O}_{2p}|\bm{\epsilon}\cdot\bm{r}|\text{O}_{1s}\rangle|^2
\end{split}
\label{eq:model_2}
\end{align}
However, using the one-body core-hole approximation, working with final-state orbitals only, we find
\begin{align}
\begin{split}
|\langle\Psi_f| \bm{\epsilon}{\cdot}\bm{R} |\Psi_i\rangle |^2
&\approx |\langle \tilde{\psi}_c|\bm{\epsilon}\cdot\bm{r}|\text{O}_{1s}\rangle|^2 \\
&= \sin^2 \theta_f |\langle \text{O}_{2p}|\bm{\epsilon}\cdot\bm{r}|\text{O}_{1s}\rangle|^2
\end{split}
\label{eq:model_3}
\end{align}
Therefore, based on the smaller value of $\theta_f$, the one-body final-state intensity is necessarily weaker than the many-electron intensity. 
The origin of this underestimation lies in erroneously formulating the excitation as a single-step transition from the core level to the final-state empty orbital,
which contains a reduced \ce{O} $2p$ component due to core-hole attraction [as illustrated in Fig. \ref{fig:results} (c)].
On the other hand, the many-electron formalism takes the correct time-ordering into account, describing a multi-step transition:
the electron is promoted to the unperturbed initial-state empty orbital followed by a many-electron charge transfer.
By this argument, the absorption intensity is the same as in the initial-state picture, $\sin^2 \theta_i |\langle \text{O}_{2p}|\bm{\epsilon}\cdot\bm{r}|\text{O}_{1s}\rangle|^2$. 
Note, however, that the energy of the \emph{final-state} configuration should be used in the Fermi's golden rule.

%For a system with more empty orbitals, we can easily extend this argument to sum over the unoccupied subspace for each final state configuration (as defined by Eq. \ref{eq:Af}).\label{key}
For the two-peak near-edge fine structure in TMOs, we can also make use of the above two-electron model.
Let us define an energy dependent hybridization within the unoccupied orbitals between metal $3d$ and \ce{O} $2p$ character according to $\sin^2\theta = \frac{t^2}{t^2 + \Delta ^ 2}$, 
where $t$ is the intrinsic hybridization strength, $\Delta(\varepsilon) = (\varepsilon + \sqrt{\varepsilon ^ 2 + t ^ 2})$, and 
$\varepsilon=\varepsilon_{3d} - \varepsilon_{2p} > 0$. Within quasi-octahedral symmetry, we would expect lower intrinsic hybridization values for the $t_{2g}$ orbitals vs. the $e_g$, but the $e_g$ orbital energies should lie above those of the $t_{2g}$. For a two-peak near-edge, we can define the peak intensities using: $t_1$ and $\varepsilon_1 = \varepsilon_{t_{2g}} - \varepsilon_{2p}$ for the lower energy $t_{2g}$ peak and $t_2$ and $\varepsilon_2 = \varepsilon_{e_g} - \varepsilon_{2p}$ for the higher energy $e_g$ peak, assuming $0 < t_1 < t_2$ and $0 <  \varepsilon_1 < \varepsilon_2$.
 
Assume, without loss of generality, that within the initial state picture the $t_{2g}$ and $e_{g}$ peaks have the same intensity: 
$\sin^2\theta_{i_1} = \frac{t_1^2}{t_1^2 + \Delta(\varepsilon_1) ^ 2} = \frac{t_2^2}{t_1^2 + \Delta(\varepsilon_2) ^ 2}=\sin^2\theta_{i_2}$.
For the purposes of illustration, we can use the following numerical values: $\varepsilon_{t_{2g}} = 1.0$, $\varepsilon_{e_g} = 4.0$, $\varepsilon_{2p} = -4.0$, and $\sin^2\theta_{i_1} = \sin^2\theta_{i_2} = 0.2$
such that $t_1 = 2.5$ and $t_2 = 4.0$ (a comparable energy unit could be eV), with the expected ordering.

If the core hole deepens the \ce{O} $2p$ orbital energy, $\varepsilon_{2p}$, to $\tilde{\varepsilon}_{2p}$, 
then the one-body final-state intensities will change and the intensity ratio decreases, as shown numerically in Table \ref{tab:sin2}.
It can be seen from this example that a one-body final-state estimate of the $3d$ peak-intensity ratio ($\sin^2\theta_{f_1} / \sin^2\theta_{f_2}$) always decreases with increasing core-hole binding.

\renewcommand{\arraystretch}{1.5}
\begin{table}
\setlength{\tabcolsep}{12pt}
\begin{tabular}{ c | c  c   c   c }
\hline
$\tilde{\varepsilon}_{2p}$		&  -4.0 ($\varepsilon_{2p}$) & -6.0	 &  -8.0 & -10.0 \\
\hline
$\sin^2\theta_{f_1}$ & 0.2 & 0.113 & 0.072 & 0.049 \\
$\sin^2\theta_{f_2}$ & 0.2 & 0.138 & 0.100 & 0.075 \\
ratio & 1.0 & 0.82 & 0.72 & 0.65 \\
\hline 
\end{tabular}
\caption{The relative near-edge peak intensities in a simple two-electron system with two available empty orbitals having \ce{O} $2p$ hybridization and energies consistent with $t_{2g}$ and $e_g$ orbitals and their dependence on the final state orbital energy $\tilde{\varepsilon}_{2p}$.}
\label{tab:sin2}
\end{table}

\subsection{Charge-transfer effects and impact on simulated spectra}
\label{sec:afi}

\begin{figure}
\centering
\includegraphics[width=0.90\linewidth]{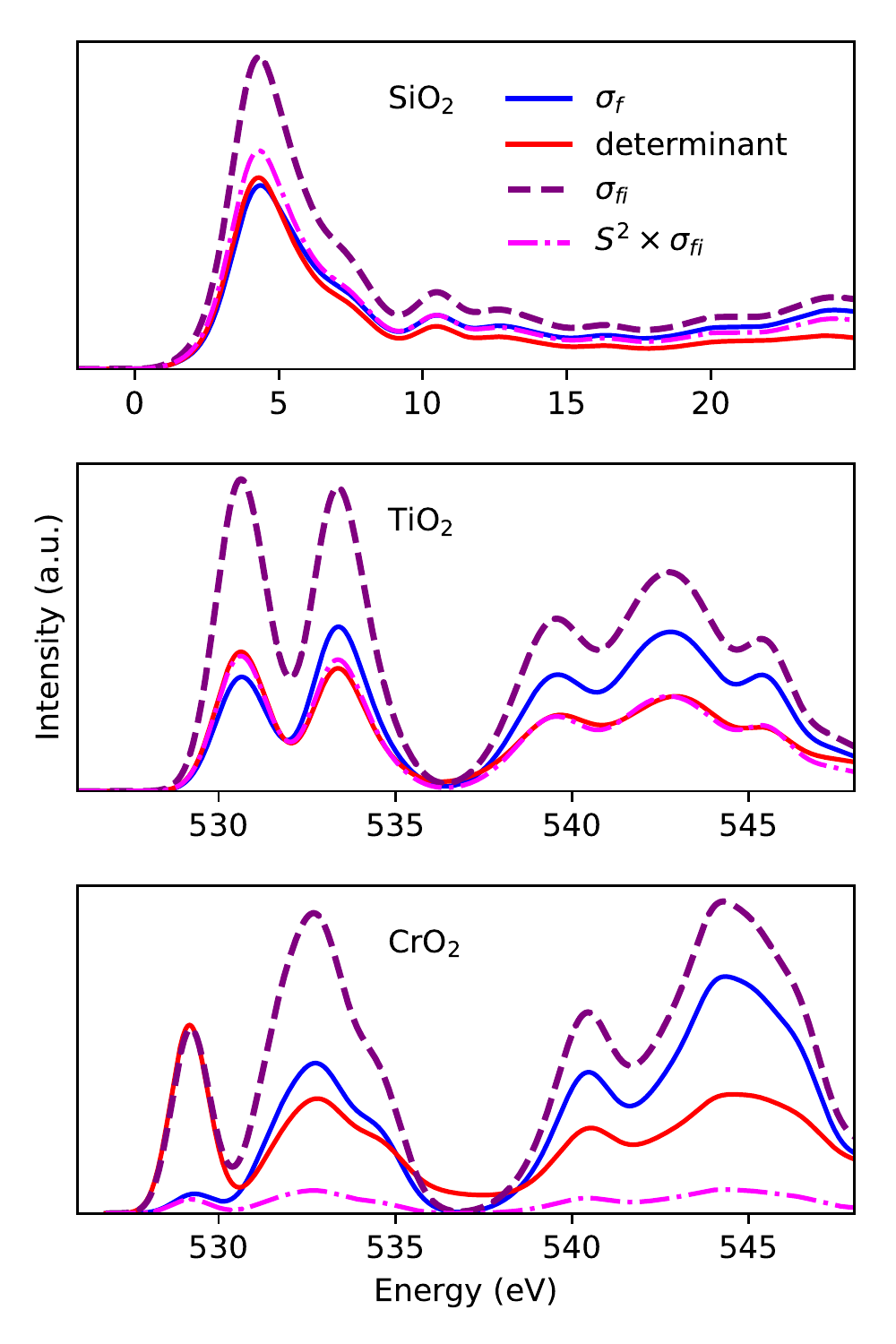}
\caption{comparison of the final-state spectra ($\sigma_i$) and the projection spectra ($\sigma_{fi}$) for \ce{TiO2}, \ce{SiO2},  and \ce{CrO2}.
The final-state spectra and the spectra from the determinantal approach (up to $f^{(2)}$) are taken from Fig. \ref{fig:results}. 
}
\label{fig:afi}
\end{figure}

While the one-body approach fails systematically in predicting the XAS for the chosen TMOs,
it produces a satisfactory lineshape for \ce{SiO2}. 
This is consistent with the previous success with using the one-body approach for a wide variety of systems \cite{taillefumier2002x, prendergast2006x, drisdell2013probing, pascal2014x, velasco2014structure, ostrom2015probing, drisdell2017determining} that are not TMOs. 
We make use of the connections between the one-body and many-body approaches outlined in Sec. (\ref{sec:onebody}) to understand why this is the case here.
 
A comparison of spectra obtained in different ways is shown in Fig. \ref{fig:afi}.
The projection spectrum is more intense than the final-state spectrum in all cases, 
indicating the hybridization term $\langle \psi_v  | \tilde{\psi}_f \rangle$ is not neligible.
However, the spectra of the chosen systems are affected in different manners by this term.
For \ce{SiO2}, the projection spectrum $\sigma_{fi}$ is in proportion to the final-state spectrum $\sigma_f$ (multiplication by the many-body overlap, $S$, correctly renormalizes the spectrum).
On the other hand, the near-edge spectral profiles in \ce{TiO2} and \ce{CrO2} are substantially modified from the one-body approximation by the projection onto empty orbitals,
in particular for \ce{CrO2} where the first peak is partly retrieved in terms of its relative intensity with respect to the second peak (around 532.5 eV).
This indicates that the projection defined in Eq. (\ref{eq:sigma_fi}) plays an important role in retrieving some key absorption features,
which makes this definition an efficient means to determine whether the final-state rule is sufficient for obtaining a satisfactory XAS.

Although the projection spectrum can rectify the deficiency of the final-state rule to some extent, 
it is still necessary to employ the determinant formalism for a correct and physical spectrum.
For \ce{CrO2}, the projection spectrum still deviates significantly from experiments, even after it is rescaled by $S$.
This suggests the many-electron effects described by the second terms in Eq. (\ref{eq:mat_decomposed}) are not trivial and should be included.

\begin{figure*}
\centering
  \includegraphics[angle=0, width=0.95\linewidth]{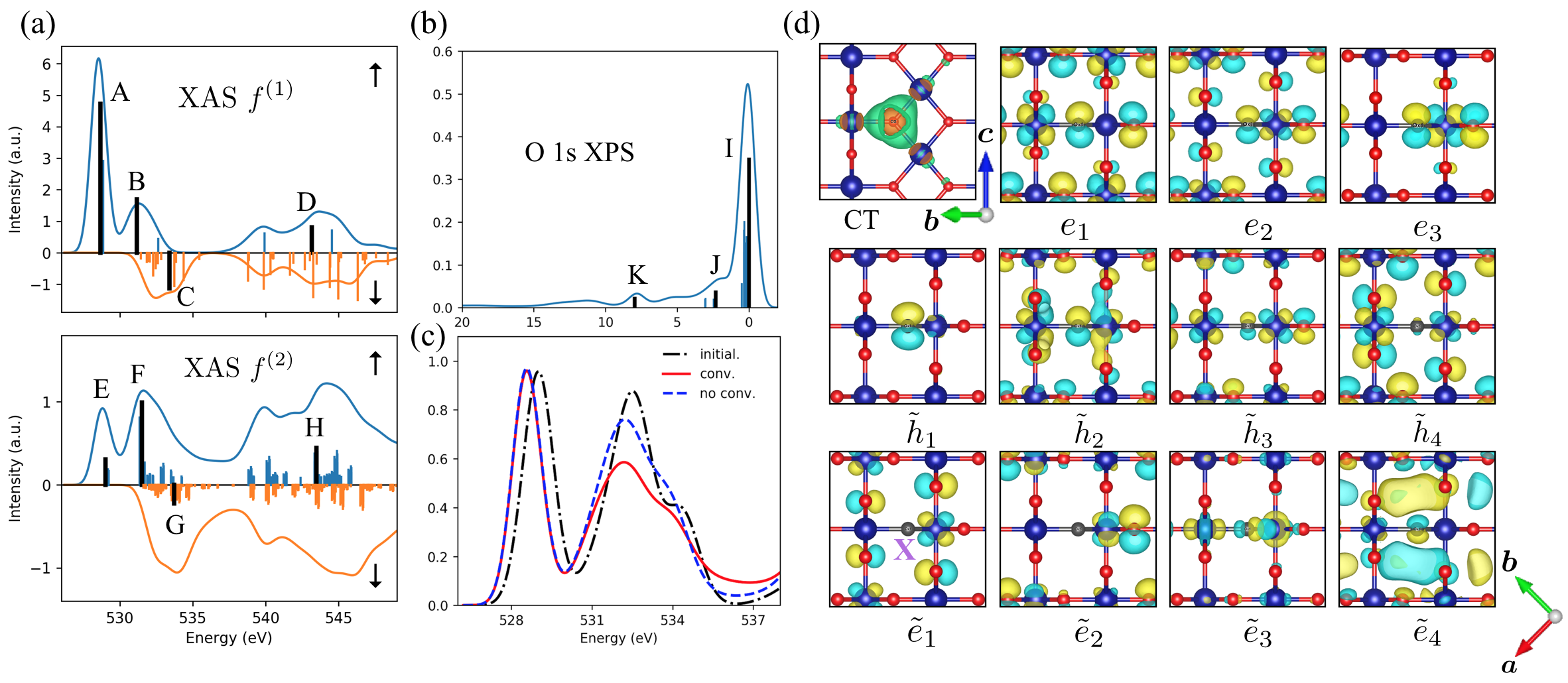}
  \caption{
 \textbf{(a)} Decomposed contributions from the single $f^{(1)}$ and double $f^{(2)}$ configurations to the \ce{O} $K$ edge XAS of \ce{CrO2}.
 For each case, the spectrum is decomposed into an spin-up ($\uparrow$) and an spin-down ($\downarrow$) channel.
 All the spectra are plotted with the same intensity scale, with sticks, i.e., oscillator strengths of the final states, in the background.
 Only $10\%$ states with the strongest oscillator strengths are shown.
 The major sticks are highlighted with black bars.
 \textbf{(b)}  \ce{O} $1s$ XPS of \ce{CrO2}. 
 The energy of the final \emph{ground-state} (with the least binding energy) is aligned with zero.
 \textbf{(c)} Comparison of the peak-intensity ratios of the initial-state spectrum ($\sigma_{A\uparrow} *\sigma_{P\downarrow} + \sigma_{A\downarrow} *\sigma_{P\uparrow}$, black), the final spectrum convoluted from the two spin-channels (red), and the fictitious spectrum without convolution ($\sigma_{A\uparrow} + \sigma_{A\downarrow}$, red). 
 The first peaks are rescaled to the same height.
 \textbf{(d)} Charge difference $\rho_f - \rho_i$ of the $N$-electron charge-transfer (CT) state and relevant 1p orbitals.
$e_i$ and $h_i$ denotes empty and occupied orbitals respectively.
Final-state orbitals are annotated with \emph{tilde}. 
$\bm{a}$ and $\bm{b}$ are two hard axes of \ce{CrO2} and $\bm{c}$ is the easy axis.
The photon polarization is in the hard-plane.
Note that the CT state is shown from a perspective different from the 1p orbitals.
For the CT plot, the charge gain (loss) is shown in orange (green).
For the orbital plots, yellow and cyan indicate the phases of the spatial wave functions.
   }
  \label{fig:cro2_details}
\end{figure*}

We consider the XAS of \ce{CrO2} in more detail. 
Fig. \ref{fig:cro2_details} (a) shows the spin-dependent $f^{(1)}$ and $f^{(2)}$ contributions to the spectrum separately, 
together with the oscillator strengths of some main transitions ($>10\%$ of the strongest transitions) presented as ``sticks''. 
We begin with an analysis of the $f^{(1)}$ terms that consist of only a single electron-core-hole pair. 
Because the core hole is fixed, an  $f^{(1)}$ term can be mapped to a single empty \emph{final-state} orbital
\begin{align}
\begin{split}
A \mapsto \tilde{e}_1 \uparrow, B \mapsto \tilde{e}_3 \uparrow, C\mapsto \tilde{e}_3\downarrow, D\mapsto \tilde{e}_4 \uparrow
\end{split}
\end{align}
where $\tilde{e}_3 \uparrow$ and $\tilde{e}_3 \downarrow$ closely resemble one another, only one of which is shown in  \ref{fig:cro2_details} (d).
The orbitals defining A, B, C, and D correspond to a $t_{2g}$ $d_{xy}$, an $e_{g}$ $d_{z^2} \uparrow $, an $e_{g}$ $d_{z^2} \downarrow$, and an unbound itinerant ($p$-like) orbital respectively.
Hereafter, $\uparrow$ is omitted unless for spin-down orbitals.

What do these transitions have in common? 
They all reflect projections of the initial (ground) state, mediated by the photon electric field, onto final states that share a common \ce{O} $1s$ core-hole excitation and its associated perturbing potential.
The core hole attracts electron density towards the excited \ce{O} site, as can be seen from the plotted isosurface of the charge-density difference $\rho_f - \rho_i$ in Fig. \ref{fig:cro2_details} (d) (top left).
This charge transfer results from the response of the $N$-electron system to the core-hole potential. 
It is computed as the deviation of final-state DFT charge density $\rho_f$ (without the excited electron as in the FCH approximation) from the one of the initial state $\rho_i$.
According to the first term in Eq. (\ref{eq:mat_decomposed}), there is a single prefactor common to all final states for this component of the $f^{(1)}$ transitions, also denoted $S$ in Eq. (\ref{eq:compare_to_f}). 
This $N$-electron determinant is yet another way of representing the CT state. 
Generally speaking, all $(N+1)$-electron final states, within this MND single-determinant picture, only differ by a few composite single-particle orbitals that slightly modulate this CT density.
The $f^{(1)}$ states differ by the addition of just one final-state unoccupied orbital.

Close examination of the final state orbitals in \ref{fig:cro2_details} (d) reveals, surprisingly, that the brightest transition of the entire spectrum originates from state $A$, 
even though its excited electron orbital, $\tilde{e}_1$ , does \emph{not overlap} with the excited \ce{O} atom [marked by \lq\lq X\rq\rq in Fig. \ref{fig:cro2_details} (d)].
%The computed projection of $e_1$ onto the $2p$ orbital of the excited \ce{O} (denoted as OX) is small:  $\langle e_1 |  \text{OX}_{2p_x}\rangle = 0.00013$.
As a result, the one-body final-state rule gives a transition amplitude of only
\begin{align}
\begin{split}
|\langle \tilde{e}_1 |x| \psi_h\rangle| = 9.16 \times10^{-6} (\text{a.u.})
\end{split}
\end{align}
which explains the lack of any significant first peak in the simulated one-body XAS of \ce{CrO2} in Figs.~\ref{fig:results} and~\ref{fig:afi}.
This small amplitude is due to Pauli-blocking resulting from the charge transfer -- in other words, the core-hole potential has lowered some initially unoccupied \ce{O} $2p$ orbital character below the Fermi level of this half-metal, rendering it inaccessible within this 1p picture.

By contrast, the many-body determinantal amplitude of $A$ is a few orders of magnitude larger:
\begin{align}
\begin{split}
 |\sum_{c\in\text{empty}} (A^f_c)^* \langle\psi_c |x| \psi_h\rangle| = 1.08\times10^{-2}
 \end{split}
\end{align}
To understand why the many-body state $A$ still has a strong oscillator strength,
an analysis can be provided based on the Laplace expansion of the determinantal amplitude in Eq. (\ref{eq:mat_decomposed}).
By inspection, we find that the most important contributions to the amplitude of $A$ are from $\tilde{\psi}_i=\tilde{e}_1$, $\tilde{h}_4$, and $\tilde{h}_3$.
They have substantial overlap (integrals tabulated in Tab. \ref{tab:A}) with a number of \emph{initial-state} empty orbitals that exhibit $p$ character at the excited \ce{O} atom, 
such as $e_1$, $e_2$, and $e_3$, as shown in Fig. \ref{fig:cro2_details} (d) (top row).
Consequently, the projection amplitudes $| \langle \tilde{\psi}_i | P_c x | \psi_h \rangle|$ of $\tilde{e}_1$, $\tilde{h}_4$, and $\tilde{h}_3$ are still significant (i.e., similar in magnitude to the amplitude of $A$), 
although these final-state orbitals may have small overlap with the core hole.
Furthermore, the corresponding many-electron overlaps, $M^f_i$, are not small (Tab. \ref{tab:A}).
Therefore, the combined contribution $\sum_i (M^f_i)^* \langle \tilde{\psi}_i | P_c x | \psi_h \rangle$ for $\tilde{\psi_i}$ in $\{ \tilde{e}_1$, $\tilde{h}_4$, $\tilde{h}_3 \}$ is significant: $6.91\times 10^{-3}$, comprising $64\%$ of the total amplitude of $A$.
From this example, it can be seen that empty initial-state orbitals and a multi-orbital picture are crucial for understanding the brightness of near-edge transitions in metallic systems.

\renewcommand{\arraystretch}{1.5}
\begin{table}
\setlength{\tabcolsep}{3.5pt}
\begin{tabular}{ c | c c c c c c}
\hline
$\tilde{\psi}_i$		
& $| \langle \tilde{\psi}_i | P_c x | \psi_h \rangle|$ 
& $ |\langle \tilde{\psi}_i | e_1 \rangle |$ 
& $ |\langle \tilde{\psi}_i | e_2 \rangle |$
& $ |\langle \tilde{\psi}_i | e_3 \rangle |$ 
& $ |M^f_{i}| $ &  \\
\hline
$\tilde{e}_1$ & $3.79\times 10^{-3}$ & $0.30$ & $0.51$ & $0.05$ & $0.34$\\
$\tilde{h}_4$ & $9.93\times 10^{-3}$ & $0.28$ & $0.08$ & $0.24$ & $0.24$\\
$\tilde{h}_3$ & $1.23\times 10^{-2}$ & $0.27$ & $0.07$ & $0.28$ & $0.26$\\
\hline 
\end{tabular}
\caption{Quantities relevant for analyzing the expansion in Eq. (\ref{eq:mat_decomposed} for state A.)}
\label{tab:A}
\end{table}

\subsection{Shake-up effects in half-metallic \ce{CrO2}}
\label{sec:more_eh}

The determinantal approach introduced in this work does not set any constraint on the number of  \ehpair{} pairs to be included
and is capable of considering more complex excitations than in the BSE.
Higher-order \ehpair{}-pair production (so-called shake-up effects due to the core-hole perturbation) should be less costly from an energy perspective in systems with smaller band gaps, and therefore more evident in the near-edge fine structure.
This section discusses these effects for the half-metallic \ce{CrO2}, whose majority-spin channel is metallic, while the minority-spin channel is insulating.
The interplay of the two spin channels in x-ray excitations gives rise to intriguing physics that cannot  be simply explained by excitonic effects. 
We will discuss how the measured XAS takes shape to illustrate additional many-body effects that are captured within the determinantal approach, beyond those already highlighted above for the  $f^{(1)}$ transitions.
%These effects include charge-transfer satellites and many-body wave function overlaps.

For \ce{CrO2}, the $f^{(2)}$ XAS contribution becomes comparable to that of $f^{(1)}$ at $\sim4.0$ eV above the absorption onset (Fig. \ref{fig:results} (a)).
The $f^{(2)}$ configurations can be considered as \emph{shake-up} excitations derived from $f^{(1)}$.
Below is the composition of some major $f^{(2)}$ configurations outlined in Fig. \ref{fig:cro2_details} (a)
\begin{align}
\begin{split}
E\mapsto(\tilde{e}_1, \tilde{h}_4, \tilde{e}_2)&, \ F\mapsto(\tilde{e}_1, \tilde{h}_3, \tilde{e}_3),\\
G\mapsto(\tilde{e}_1, \tilde{h}_3, \tilde{e}_3\downarrow)&, \ H\mapsto(\tilde{e}_1, \tilde{h}_3, \tilde{e}_4)
\end{split}
\end{align}
They can be derived from the $f^{(1)}$ states by adding one more \ehpair{} pair
\begin{align}
\begin{split}
E\mapsto A + (\tilde{e}_2, \tilde{h}_4)&, \ F\mapsto B + (\tilde{e}_1, \tilde{h}_3), \\
G\mapsto C + (\tilde{e}_1,\tilde{h}_3)&, \ H\mapsto D + (\tilde{e}_1, \tilde{h}_3)
\end{split}
\end{align}
where $\tilde{h}_3$, $\tilde{h}_4$, $\tilde{e}_1$, and $\tilde{e}_2$ are $t_{2g}$ orbitals close to the Fermi level.
As is shown in Fig. \ref{fig:cro2_details} (d), 
orbital $\tilde{e}_1$ has significant spatial overlap with $\tilde{h}_3$ (sharing the $d_{xz}$ character at the \ce{Cr} atom next to the excited \ce{O}),
and so does orbital $\tilde{e}_2$ with $\tilde{h}_4$ (near the oxygens at the corners of the plot), albeit weaker.
This overlap makes $E$, $F$, $G$, and $H$ also bright transitions.
There are alternative pathways to access these states with two \ehpair{} pairs.
For instant, $F$ can also be mapped to $A + (\tilde{e}_3, \tilde{h}_3)$, i.e., $A$ coupled with an \ehpair{} pair $(\tilde{e}_3, \tilde{h}_3)$ (a shake-up $d-d$ transition).

The shake-up excitations can also be found in the satellite features of XPS, as shown in Fig. \ref{fig:cro2_details} (b).
Recently these excitations were investigated with a cumulant expansion technique \cite{kas2015real, kas2016particle}.
Here, we show that these satellite features can also be included naturally within the determinant formalism of the non-interacting MND theory (albeit poorly approximating their energies due to missing additional interactions between these extra \ehpair{} pairs).
The strongest transition (labelled as state $I$) originates from the overlap of the $N$-electron states, describing the initial ground state valence system and the final core-excited valence system (assuming the excited electron has escaped, approximated using the full-core-hole approach): $\langle \Psi^N_{f, \text{FCH}} | \Psi^N_{i, \text{GS}} \rangle$.
This corresponds to the charge-transfer state in Fig. \ref{fig:cro2_details} (d). 
We may define $I$ as the only zero-order configuration ($f^{(0)}$) of XPS.
$f^{(1)}$ configurations emerge at larger binding energies and appear as satellite features in the XPS profile.
Two representative states are $J$ and $K$
\begin{align}
\begin{split}
J\mapsto(\tilde{e}_1, \tilde{h}_2), \ K\mapsto(\tilde{e}_1, \tilde{h}_1)
\end{split}
\end{align}
which are shake-up excitations from $\tilde{h}_2$ (a  \ce{Cr} $3d$ - \ce{O} $2p$ hybrid with mixed bonding and anti-bonding character) and $\tilde{h}_1$ (a deep \ce{O} $2p$ orbital) to the $\tilde{e}_1$ orbital, respectively.
The charge transfer associated with $K$ is particularly strong.

\subsection{Many-body wavefunction overlap effects in \ce{CrO2}}
\label{sec:spin_conv}

As shown in Sec. \ref{sec:afi},  the projection onto empty initial-state orbitals alone cannot account for the XAS lineshape for \ce{CrO2},
and one must employ the determinant formalism.
This suggests that there are important many-electron effects in the determinantal amplitude that lead to the ultimate peak-intensity ratio of $\sim1.7$ between the first and second absorption features.
To explain this, we rewrite the spectrum as the convolution defined in Eq. (\ref{eq:xas_conv})
\begin{align}
\begin{split}
\sigma_{A}&=\sigma_{A\uparrow} *\sigma_{P\downarrow} + \sigma_{A\downarrow} *\sigma_{P\uparrow} 
\end{split}
\end{align}
where 
$\sigma_{A}\equiv\sigma_\text{XAS}, \sigma_{P}\equiv\sigma_\text{XPS}$, 
$*$ represents the convolution integral in Eq. (\ref{eq:xas_conv}), 
$\sigma_{A\mu}$ and $\sigma_{P\mu}$ are spectra of one-spin channel before convolution.
Then the spectral functions $\sigma_{P\mu}$ can be considered as weighting factors of the two absorption channels $\sigma_{A\mu}$. 
If the weighting factors are not considered, the hypothetical spectrum
\begin{align}
\begin{split}
\sigma'_{A}&=\sigma_{A\uparrow}  + \sigma_{A\downarrow}  
\end{split}
\end{align}
has a peak-intensity ratio of $\sim 1.3$ that still deviates significantly from experiment (Fig. \ref{fig:cro2_details} (c)).
This implies that the modulation effects of $\sigma_{P\uparrow}$ and $\sigma_{P\downarrow}$ on their counter-spin channel are quite different.

\begin{figure}
\centering
  \includegraphics[angle=0, width=0.95\linewidth]{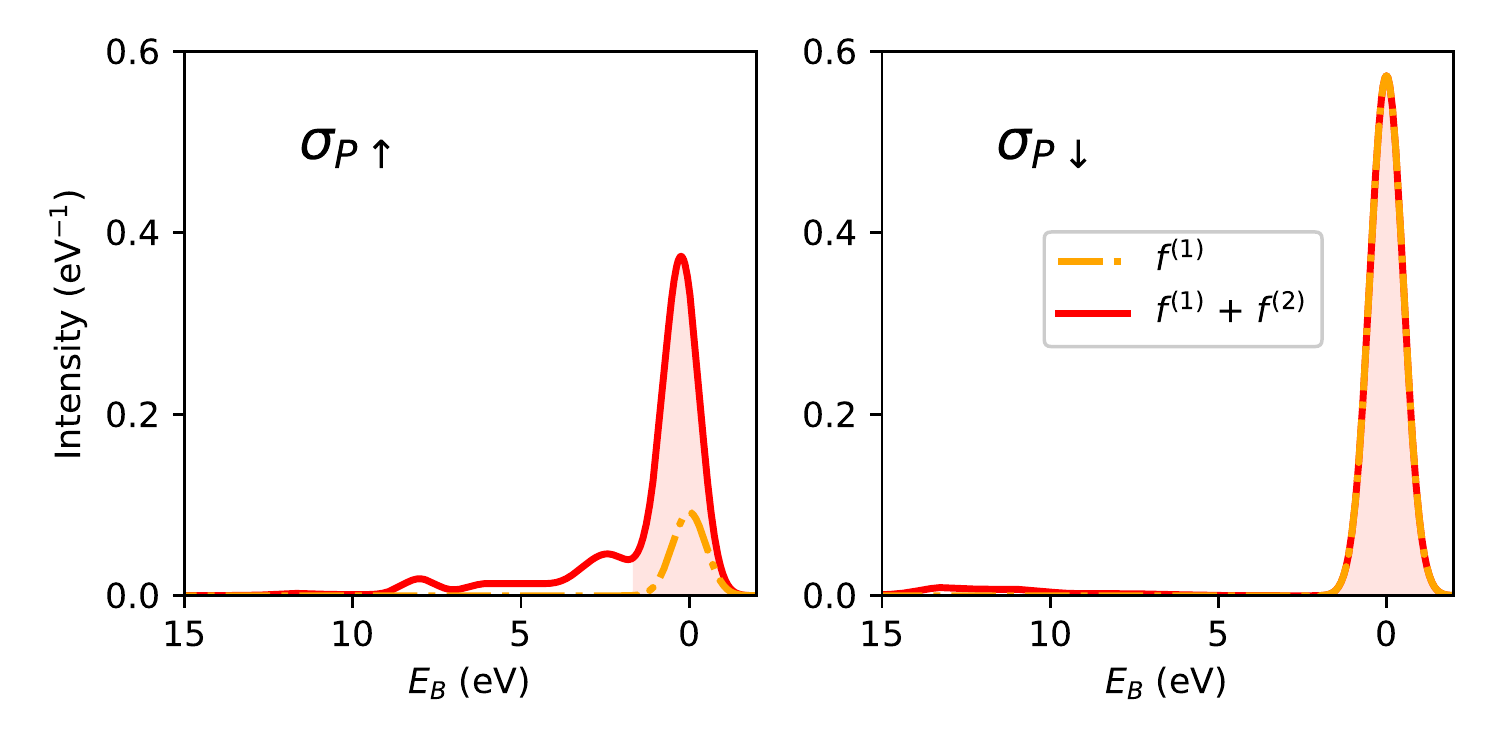}
  \caption{Spin-wise spectral function $\sigma_{P\uparrow}$ and $\sigma_{P\downarrow}$.
  The x-axis is the binding energy ($E_B$).
  $E_B = 0$ is aligned with the threshold.
   }
  \label{fig:cro2_xps}
\end{figure}

The spectral functions, $\sigma_{P\uparrow}$ and $\sigma_{P\downarrow}$, are shown in Fig. \ref{fig:cro2_xps}.
In both cases, most spectral weight is concentrated at zero binding energy, $E_B = 0$.
But for the metallic $\uparrow$ channel, more spectral weight is transferred to shake-up satellites at higher energies
because its lack of a band gap makes \ehpair{} pair production easier.
As a result, $\sigma_{P\downarrow}$ is more intense than $\sigma_{P\uparrow}$ near $E_B = 0$.
The integrated intensity of $\sigma_{P\uparrow}$ is $\sim 2/3$ of $\sigma_{P\downarrow}$ for $E_B < 1.7$ eV (shaded areas).
The more intense $\sigma_{P\downarrow}$ enhances the contribution of $\sigma_{A\uparrow}$,
especially the lowest-energy peak defined by $t_{2g}\uparrow$ orbitals,
leading to a peak-intensity ratio of $\sim 1.7$ as measured.

To conclude, the three contributing factors leading to the near-edge lineshape of \ce{CrO2} are:
(a) the core-level excitonic effect in the metallic screening environment lead to a mild increase in the edge intensity 
(the initial-state spectrum is also shown in Fig. \ref{fig:cro2_details});
(b) shakeup excitations in the spin-up channel reduces the many-body wave function overlap $\sigma_{P\uparrow}$ at $E_B=0$;
(c) the smaller wave function overlap (orthogonality effects) reduces the intensity of the spin-down channel that mainly contributes to the second absorption feature, 
leading to a even stronger first peak versus the second.

\section{Numerical considerations and computational efficiency}
\label{sec:comp}

\begin{figure*}
\centering
  \includegraphics[angle=0, width=0.98\linewidth]{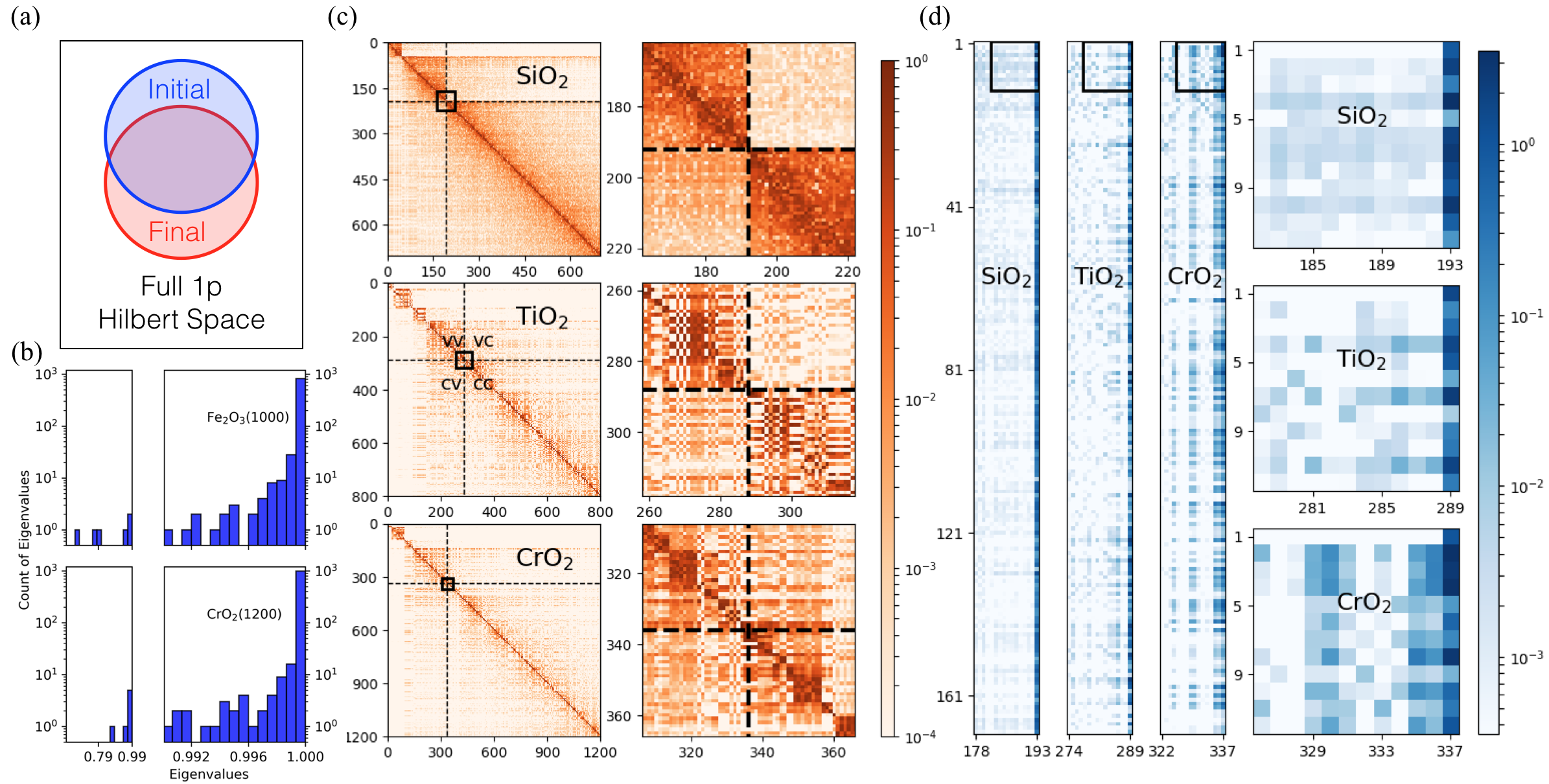}
  \caption{
 \textbf{(a)} 
 Schematic showing the relation of the subset of initial(final)-state orbitals chosen in practical calculations to the full Hilbert space.
 \textbf{(b)} 
 Histograms for the distribution of the eigenvalues of the square $\bm{\xi}'s$. 
 Counts of eigenvalues are in logarithm scale.
 The bar widths (above $0.9$) are $\frac{2}{3}\times10^{-3}$.
 \textbf{(c)} 
 $\xi$-matrices for \ce{SiO2}, \ce{TiO2}, and \ce{CrO2} $\uparrow$.
 The dashed lines mark the Fermi level of the initial (vertical) and final (horizontal) state.
 The right panels are the regions enclosed by the bolded squares on the left ones near the crossings of the two Fermi levels.
 Within these regions, the \ce{CrO2} has large matrix elements in all four quadrants while the large matrix elements are mainly located within the $vv-$ or $cc-$block for \ce{SiO2} and \ce{TiO2}. 
 \textbf{(d)} 
 $\zeta$-matrices that correspond to the $\xi$-matrices in (c).
 Rows iterates over empty-orbital indices with $1$ being the lowest empty one.
 Columns iterates over occupied-orbital indices with $1$ being the lowest occupied one.
 Right panels are enlarged views of the square regions in the left ones.
 Both (c) and (d) display the absolute values of the complex matrix elements in logarithm scale.
   }
  \label{fig:numerics}
\end{figure*}

\subsection{Properties of the $\xi$-matrix}
\label{sec:xi}

One primary concern of the determinantal approach is the numerical accuracy of the $\xi$-matrix ($\bm{\xi}$).
In practice, one can only choose a finite number of orbitals (bands) in first-principles calculations and 
this set of orbitals can not span the full 1p Hilbert space, as illustrated in Fig. \ref{fig:numerics} (a).
Therefore, the initial-state orbital set may not overlap with the final-state one, 
resulting in a $\bm{\xi}$ that is projective rather than unitary.
Furthermore, it may be worrisome if the numerical error in the matrix elements of $\bm{\xi}$ is accumulative, leading to determinant values that are either vanishingly small or unrealistically large. 

Here, we demonstrate that using the optimal basis set for expanding 1p wave functions can produce a $\xi_{ij}$ matrix close to unitary,
such that the spectral weight of the determinantal spectrum is on the same order of magnitude as the 1p final-state spectrum as compared in Sec. \ref{sec:afi}.
When constructing the Shirley optimal basis sets, we include a sufficient number of bands (Tab. \ref{tab:comp}) so that 
the optimal basis functions can cover a range of 1p wave functions, from localized $3d$-orbitals to delocalized states.
We measure the quality of a transformation matrix by its eigenvalues.
A close-to-unitary transformation matrix should have eigenvalues that are close to $1$ predominantly.
Through examining $\bm{\xi}$ of the studied systems,
we find more than $90\%$ of the eigenvalues are larger than $0.995$, with a maximum below $1.0001$, 
which suggests these $\bm{\xi}$'s are close to unitary.
A typical statistics of the eigenvalues of $\bm{\xi}$ using \ce{Fe2O3} and \ce{CrO2} $\uparrow$ as examples is provided in Fig. \ref{fig:numerics} (b).

The second concern regarding the practicality of the determinantal approach is how many configurations are relevant for a converged lineshape.
From the analysis of the BFS algorithm, we know that this depends on the sparsity of $\bm{\zeta}$ and how many non-vanishing minors one can extract from $\bm{\zeta}$.

We first analyze the properties of $\bm{\xi}$.
Fig. \ref{fig:numerics} (c) displays the $\bm{\xi}$ for the three representative cases, 
the large-band-gap \ce{SiO2} ($690\times690$),
the semiconducting \ce{TiO2} ($800\times800$),
 and the metallic spin channel ($\uparrow$) of \ce{CrO2} ($1200\times1200$).
All $\bm{\xi}$'s are \emph{quasi-block-diagonal}, which indicates the core-hole-induced hybridization mainly occurs within orbitals of similar energies.
Overall, the $\bm{\xi}$ of \ce{SiO2} and \ce{TiO2} has more off-diagonal matrix elements compared to \ce{CrO2}
because electronic screening of the core hole is weaker in an insulator/semiconductor than in a metal.
In the region near the Fermi levels, however, the $\bm{\xi}$ of \ce{SiO2} and \ce{TiO2} has less off-diagonal matrix elements than \ce{CrO2}: 
for \ce{SiO2} and \ce{TiO2}, the significant matrix elements are mainly concentrated at the vv-(occupied-to-occupied) and cc-(empty-to-empty) blocks; 
but for \ce{CrO2}, there are more non-vanishing matrix elements in the vc- or cv-block, especially in the vicinity of the Fermi-level crossing.
This is because the Fermi surface of a metallic system is susceptible to the core-hole potential, which strongly rehybridizes the orbitals near the Fermi surface.

$\bm{\xi}$ of \ce{SiO2} is also significantly different from those of \ce{TiO2} and \ce{CrO2}.
The distribution of nontrivial matrix elements is more homogeneous within the $vv$ and $cc$ block for \ce{SiO2} compared to \ce{TiO2} or \ce{CrO2}.
This is also consistent with the analysis with projection spectra in Sec. \ref{sec:afi}:
the conduction bands of \ce{SiO2} hybridize uniformly with the valence bands due to the core hole,
leading to very similar lineshapes in the 1p, projection, and determinantal spectrum,
whereas the $cv$-hybridization in \ce{TiO2} or \ce{CrO2} is less uniform and orbital-dependent, 
leading to a few-body molecular description of x-ray excitations as in Sec. \ref{sec:afi} and \ref{sec:more_eh}.

\renewcommand{\arraystretch}{1.5}
\begin{table*}
\begin{tabular}{ c | c c | c c | c c c c}
\hline
\hline
System 						& Calc. $E_g$ (eV)	& $|\det|$			& \#Orb.	& \#Elec.		& \# $f^{(2)}$ (M)	& \# Significant $f^{(2)}$ (M) 	& Ratio ($\%$) \\
\hline
 \ce{TiO2}						& 1.79		& 0.8077	&  800 			& 288				& 37.7				& 0.0567							& 0.15\\			
 \ce{SrTiO3}					& 2.26		& 0.8125	& 1200			& 540				& 117				& 0.891							& 0.76\\  
 \ce{Fe2O3}					& 1.10		& 0.7922	& 1000			& 400				& 71.9				& 0.197							& 0.27\\
 \ce{VO2}						& 0.00      & 0.7594	& 800				& 300				& 37.4				& 0.0345							& 0.09\\
 \ce{CrO2$\uparrow$}		& 0.00		& 0.3397	& 1200			& 336				& 125				& 0.242							& 0.19\\
 \ce{CrO2$\downarrow$}	& 3.68		& 0.8474	& 1200			& 288				& 119				& 0.282							& 0.24\\
 \ce{MnO2}					& 0.09		& 0.8047	& 800				& 324				& 36.6				& 0.708							& 1.9\\
 \ce{NiO} 						& 3.33		& 0.8122	& 500				& 256				& 7.59				& 0.0812	 						& 1.1\\
 \ce{CuO}						& 0.12		& 0.4871	& 1024			& 544				& 62.5				& 0.604							& 0.97\\
 %\ce{Diamond} 				& 3.96		& 0.9332	& 1100			& 432				& 96.2				& 0.0782							& 0.08\\
 \ce{SiO2} 				& 6.19		& 0.8370	& 690			& 192				& 23.7				& 0.179							& 0.75\\
\hline
\hline 
\end{tabular}
\caption{ 
(Initial-state) band gaps $E_g$ obtained on the DFT ($+U$) level; 
the absolute values of the determinants for the transformation matrix from the initial to final state for the $N$-electron systems, i.e., $|\langle\Psi^N_i | \Psi^N_f \rangle|(|\det|)$, of individual spin channels without the photoelectron; 
numbers of all orbitals and those of the occupied ones;
numbers of all $f^{(2)}$ configurations and the prominent ones that contribute to converged lineshapes and their proportions among the whole;
for the systems being studied;
}
\label{tab:comp}
\end{table*}

\subsection{Properties of the $\zeta$-matrix}
\label{sec:zeta}

Consider an ideal situation where there is no hybridization induced between the occupied and empty orbitals as the core-hole potential is introduced.
The $\xi$-matrix is exactly block diagonal and $\bm{\zeta}$ only has non-zero matrix elements in its last column.
The actual $\zeta$-matrix can be considered as a deviation from this ideal situation.
How much it deviates depends on the hybridization of the occupied and empty orbitals.
Fig. \ref{fig:numerics} (d) displays the $\zeta$-matrices for \ce{SiO2}, \ce{TiO2}, and \ce{CrO2} $\uparrow$, 
for the region that spans the lowest 170 unoccupied orbitals (rows) and the topmost 16 occupied orbitals plus the lowest unoccupied orbital (columns).
Near the Fermi levels, the $\xi$-matrices of \ce{SiO2} and \ce{TiO2} are quasi-block-diagonal, which leads to a $\zeta$-matrix with significant matrix elements mainly located on its last column. 
There are relatively a small number of non-vanishing $2\times2$ or high-order minors, and
therefore the XAS converges mostly at the $f^{(1)}$ order. We can also see that the more uniform, reduced coupling between occupied and unoccupied orbitals in \ce{SiO2} leads to a $\zeta$-matrix with a more dominant final column. By contrast, the hybridization across the band gap in \ce{TIO2} exhibits less uniformity, reflecting the existence of more localized orbitals subspaces affected by the core-hole potential, and the corresponding $\zeta$-matrix exhibits more significant terms outside the final column, indicating that the many-body approach may be more accurate for \ce{TiO2}.
For \ce{CrO2} with strong hybridization, $\bm{\zeta}$ has more significant matrix elements beyond the last column.
These matrix elements form several strips with widths of a few columns, leading to more nontrivial high-order minors.

\subsection{Computational overhead}
\label{sec:overhead}

The computational complexity of the BFS depends on how many nontrivial minors can be found from $\bm{\zeta}$.
A statistics of the computational effort required to converge XAS is shown in Table. \ref{tab:comp}.
The XAS is simulated with a supercell with dimensions around 10 \AA{} and several hundred ($N_v$) electrons.
To cover an energy window up to $20$ eV above onset, another few hundred ($N_c$) empty orbitals are also included.
Since the investigated XAS converges at the $f^{(2)}$ order,  we use the number of nontrivial $f^{(2)}$ configurations as a measure of the computational costs.
There are $N_c (N_c - 1) N_v / 2$ $f^{(2)}$ configurations in total,
whose numbers are from tens to hundreds of millions for the investigated systems.
The number of the nontrivial $f^{(2)}$ configurations as found by the BFS algorithm is typically around $1\%$ of the total.
In all of the investigated systems, this translates to at least a 100-fold speed-up of calculations, thanks to the BFS algorithm that screens out configurations of weak transition amplitudes.
For insulators such as diamond or \ce{TiO2}, even fewer configurations are needed to achieve convergence. 
The overall trend for the computational cost is: the smaller the band gap ($E_g$), the more the valence orbitals tend to hybridize with the empty orbitals (due to the core-hole potential), the smaller the determinant of the overlap matrix between the initial- and final-state ($\langle\Psi^N_i | \Psi^N_f \rangle$), and more configurations and computational efforts are required.

\section{Conclusions and Outlooks}
In conclusion, we have implemented an efficient algorithm for simulating x-ray absorption spectra (XAS) employing transition amplitudes computed within a many-body determinantal ansatz.
The core of the algorithm exploits the linear dependence of the determinants representing various electronic configurations for a fixed number of electrons and 
a breadth-first search (BFS) graph algorithm that efficiently and controllably neglects configurations whose contributions are insignificant to computed XAS, as defined by some numerical tolerance.
The new methodology has been applied to study a series of transition metal oxides (TMOs), and this simulation technique can be readily used for interpreting XAS of these technologically important materials. In the majority of cases, this approach provides an accuracy comparable to or exceeding Bethe-Salpeter equation (BSE) solutions and naturally includes electronic confgurations representing higher-order excitations beyond the subset of Feynman diagrams accessible within the BSE.

The determinantal approach can be extended to other types of x-ray spectra besides XAS, such as X-ray photoemission spectroscopy (XPS) and resonant inelastic x-ray scattering (RIXS), 
using a similar linear algebra technique and search algorithm.
It will be worthwhile to compare this new method with recent studies that apply a cumulant expansion to capture the charge-transfer satellites in XPS \cite{kas2015real, kas2016particle, zhou2017cumulant}.
And it will be interesting to test the efficiency of the current approach to produce 2D RIXS spectra that provide rich information for materials characterization.

The main drawback of the current approach relates to its approximation of the various final state configurations, which are currently derived from a single (core-orbital excited-state) self-consistent field and its associated valence Kohn-Sham orbitals. The spectrum of excitation energies within this orbital space neglects additional valence-orbital excited-state electron interactions. Therefore it cannot describe further excitonic final-state effects resulting from the shake-up of additional valence electron-hole pairs nor coupling with many-body collective modes, such as  plasmon excitations. These effects can be captured within the cumulant expansion through accurate determination of the valence dielectric response function beyond the random-phase approximation.
However, this is an excellent approximation for higher-order contributions to the spectra of metallic or semi-metallic systems, as demonstrated here for CrO$_2$, and future work will explore solutions for an interacting picture to refine our description of higher-order excited states of semiconductors and insulators and their associated spectral features.

\label{sec:outlook}

%%%%%%%%%%%%%%%%%%%%%%%%%%%%%%%%%%%%%%%%%%%%%%%%%%%%%%%%%%%%%%%%%%%%%%%%%%%%%%%%%%%%%%%%%%%%
% Appendix
%%%%%%%%%%%%%%%%%%%%%%%%%%%%%%%%%%%%%%%%%%%%%%%%%%%%%%%%%%%%%%%%%%%%%%%%%%%%%%%%%%%%%%%%%%%%

\begin{appendix}
\section{PAW formalism for obtaining the overlap matrix elements}
\label{sec:paw}
To obtain the transition amplitude $A^f_c$, a prerequisite is to find the overlap integral between the initial- and final-state Kohn-Sham orbitals, i.e., the matrix elements $\xi_{ij} = \langle\psi_j |\tilde{\psi}_i \rangle$.
In our implementation of the $\Delta$SCF calculations, we employ a plane-wave basis and the electron-ion interaction is modeled using Vanderbilt's ultrasoft pseudopotentials.
The computational efficiency gain through the use of a smaller plane-wave energy cutoff compared to what might be required when using norm-conserving pseudopotentials is offset by some additional steps in the formalism which account for using non-orthogonal projections in the pseudopotential. 
In the above calculations with the many-electron method, we have used the PAW formalism to find the overlap matrix elements $\xi_{ij}$ and here we provide the details for finding these quantities.

In the PAW formalism, the real (all-electron, AE) wave function is reconstructed from the pseudo (PS) wave function via a linear transformation $\mathcal{T}$
\begin{align}
\begin{split}
|\psi^{\text{AE}} \rangle= \mathcal{T}|\psi^\text{PS}\rangle
\end{split}
\end{align}
In practice, there is one such $\mathcal{T}$ for each pseudized atom. To simplify notation, we will omit the sum over atomic indices, $I$, for most of what follows, until it is necessary to the discussion.
$\mathcal{T}$ is responsible for correcting the wave function within the augmented spherical region $\Omega$ centered at the atom of interest.
First, $\mathcal{T}$ projects the pseudo wave function onto the preselected projectors $|p_l\rangle$ of a particular angular momentum $l$; 
then $\mathcal{T}$ corrects the wave function in the augmented region using the difference of the real and pseudo atomic wave functions of the corresponding $l$, i.e., $|\phi_l^\text{AE}\rangle - |\phi_l^\text{PS}\rangle$, and scales the wave function difference with the projection amplitude. 
$|\phi_l^\text{AE}\rangle$ and $|\phi_l^\text{PS}\rangle$ and the associated projectors are all determined when generating the pseudopotential.
Put together, the linear transformation reads
\begin{align}
\begin{split}
\mathcal{T}=1 + \sum_{l}(|\phi_l\rangle-|\phi^\text{PS}_l\rangle)\langle p_l | 
\end{split}
\end{align}
For simplicity, \quotes{AE} is dropped and only \quotes{PS} is kept to indicate a wave function is pseudo. 

A PAW construction satisfies the following conditions: 
(i) the projector functions $\langle \bm{r} | p_l \rangle$ are zero outside the augmented region $\Omega$; 
(ii) the difference of the atomic wave functions of the same $l$, i.e., $\langle \bm{r} | \phi_l \rangle - \langle \bm{r} | \phi^\text{PS}_l \rangle$, is also zero outside $\Omega$;
(iii) and we have an orthogonality and completeness relation: $\langle \phi^\text{PS}_i | p_j \rangle = \delta_{ij}$, for all $i$ and $j$, and $ P=\sum_l | \phi^\text{PS}_l\rangle \langle p_l| $ is the identity operator over $\Omega$. 
It should be noted that in the PAW formalism each angular momentum may have more than one channel so $\langle \phi_l | \tilde{\phi}_{l'} \rangle = \delta_{ll'}$ may not hold in general.

As stated, there is one such linear transformation $\mathcal{T}$ for each type of atom (i.e., for each element) and the projections should include a structure (phase) factor to account for different atomic positions within an extended, periodic context.
In the x-ray core-hole approach, however, we introduce a new type of atom. We have, as before, the initial-state (ground-state) atoms and one new type to describe the final-state atom with an excited core hole. 
In practice, this means that there are two sets of projectors and atomic wave functions involved for this particular atom. 
If one wants to obtain the overlap matrix elements $\xi_{ij}$, it is necessary to obtain overlap integrals of two wave functions that are reconstructed from two different PAW constructions.
Here, we focus on the single-atom case and find the expression for the overlap.
Consistent with the notation in the rest of the manuscript, we use a tilde to denote quantities related to the final (excited) state.  
Omitting the irrelevant indices, the overlap between an initial-state and a final-state orbital is
\begin{align}
\begin{split}
\langle \psi | \tilde{\psi} \rangle = \langle \psi^\text{PS} | \mathcal{T}^\dagger \tilde{\mathcal{T}} | \tilde{\psi}^\text{PS} \rangle
\end{split}
\label{eq:psipsi}
\end{align}
where $\psi$ and $\tilde{\psi}$ are reconstructed from two different linear transformations, $\mathcal{T}$ and $\tilde{\mathcal{T}}$.
Expanding the operator product, we find
\begin{align}
\begin{split}
&\mathcal{T}^\dagger \tilde{\mathcal{T}} \\
&= 1 
+ \sum_{l} | p_l \rangle ( \langle \phi_l | - \langle \phi^\text{PS}_l | ) 
+ \sum_{l'} (| \tilde{\phi}_{l'} \rangle - | \tilde{\phi} ^ \text{PS}_{l'} \rangle)\langle \tilde{p}_{l'} | \\
&+ \sum_{ll'} | p_l \rangle ( \langle \phi_l | - \langle \phi^\text{PS}_l | ) (| \tilde{\phi}_{l'} \rangle - | \tilde{\phi} ^ \text{PS}_{l'} \rangle)\langle \tilde{p}_{l'} |
\end{split}
\label{eq:tt}
\end{align}
This expansion can be regrouped and simplified by making use of the properties of the projectors and PAW atomic wave functions in conjunction with the completeness relation. 

First, the last summation in Eq. (\ref{eq:tt}) can be further expanded so as to make use of the projection operators $ P=\sum_l | \phi^\text{PS}_l\rangle \langle p_l| $ and $\tilde{P}$, as follows:
\begin{align}
\begin{split}
&\sum_{ll'} | p_l \rangle    ( \langle \phi_l | - \langle \phi^\text{PS}_l | )    (| \tilde{\phi}_{l'} \rangle - | \tilde{\phi} ^ \text{PS}_{l'} \rangle)   \langle \tilde{p}_{l'} | \\
=&\sum_{ll'}  | p_l \rangle    ( \langle \phi_l | \tilde{\phi}_{l'} \rangle -  \langle \phi^\text{PS}_l | \tilde{\phi} ^ \text{PS}_{l'} \rangle )    \langle \tilde{p}_{l'} | \\
- &\sum_{l} | p_l \rangle ( \langle \phi_l | - \langle \phi^\text{PS}_l | )    \big( \sum_{l'} | \tilde{\phi} ^ \text{PS}_{l'} \rangle \langle \tilde{p}_{l'} |  \big) \\
- &\big( \sum_{l} | p_l \rangle \langle \phi^\text{PS}_l | \big)    \sum_{l'} (| \tilde{\phi}_{l'} \rangle - | \tilde{\phi} ^ \text{PS}_{l'} \rangle)\langle \tilde{p}_{l'} |
\end{split}
\label{eq:last_sum}
\end{align}
The last two terms can be regrouped with the two single-summations over $l$ and $l'$ in Eq. (\ref{eq:tt}). 
For instance, the second summation in Eq. (\ref{eq:last_sum}) can be combined with the second term in Eq. (\ref{eq:tt}) as
\begin{align}
\begin{split}
\sum_{l} | p_l \rangle ( \langle \phi_l | - \langle \phi^\text{PS}_l | ) (1 - \tilde{P})
\end{split}
\label{eq:proj}
\end{align}
While the operator $| p_l \rangle ( \langle \phi_l | - \langle \phi^\text{PS}_l | ) $ is only non-zero within $\Omega$,  $(1 - \tilde{P})$ projects the wave function onto the region complementary to $\tilde{\Omega}$.
Therefore, we can consider the union of the augmented regions for the ground-state and the core-excited atom, $\Omega \cup \tilde{\Omega}$ as a volume within which the product of these operators will zero out any wave function, and the operator in Eq. (\ref{eq:proj}) is a zero operator.
In practice, we can set the radial limit for atomic integrals, like $\langle \phi | \tilde{\phi} \rangle$ to the maximum of the cutoff radii used when generating the pseudopotentials for the ground-state and the core-excited atom. More often than not, these cut-off radii are identical and $\Omega = \tilde{\Omega}$.
And so, the second and third summations in Eq. (\ref{eq:last_sum}) cancel exactly with the second and third terms in Eq. (\ref{eq:tt}).

With all terms combined, the final expression for the operator product is simplified as
\begin{align}
\begin{split}
\mathcal{T}^\dagger \tilde{\mathcal{T}} = 1 + \sum_{ll'}  | p_l \rangle    ( \langle \phi_l | \tilde{\phi}_{l'} \rangle -  \langle \phi^\text{PS}_l | \tilde{\phi} ^ \text{PS}_{l'} \rangle )    \langle \tilde{p}_{l'} |
\end{split}
\label{eq:tt_final}
\end{align}
In a multi-atomic system, the overlap matrix elements in Eq. (\ref{eq:psipsi}) can be written as
\begin{align}
\begin{split}
\langle \psi | \tilde{\psi} \rangle 
&= \langle \psi^\text{PS} | \tilde{\psi}^\text{PS} \rangle \\
&+ \sum_{I, ll'}  \langle \psi^\text{PS} |  p^I_l \rangle    ( \langle \phi^I_l | \tilde{\phi}^I_{l'} \rangle -  \langle \phi^{I,\text{PS}}_l | \tilde{\phi} ^ {I,\text{PS}}_{l'} \rangle )    \langle \tilde{p}^I_{l'} | \tilde{\psi}^\text{PS} \rangle 
\end{split}
\label{eq:psipsi_final}
\end{align}
in which the index $I$ goes over all the PAW atoms in the system.
Here, we only consider one core-excited atom within a given supercell, and so, for all but one of the atoms, the initial and final state PAW projections are identical (i.e., we can drop the tildes).

The first term in Eq. (\ref{eq:psipsi_final}) can be obtained efficiently using the pseudo wave functions in their native plane-wave basis.
The routines to evaluate the projection amplitude $\langle \psi^\text{PS} |  p^I_l \rangle$ are already required to obtain the core-level position matrix operator at the core-excited atom (as we have done in the past for the one-body final state approach).
The same procedure can be trivially extended to estimate projection amplitudes for all atoms and for both the initial and final state, using outputs from the pseudopotential generation.
An additional routine is needed for the atomic overlap term $S^I_{ll'} \equiv \langle \phi^I_l | \tilde{\phi}^I_{l'} \rangle -  \langle \phi^{I,\text{PS}}_l | \tilde{\phi} ^ {I,\text{PS}}_{l'} \rangle$, 
which can be obtained beforehand using the atomic wave functions from two given PAW constructions. All of these quantities can be computed and stored in advance for an established set of pseudopotentials and then used for any number of further periodic calculations.

\section{Optimal Basis Set for Obtaining Electronic Structure over Dense $\bm{k}$-Grid}
\label{sec:obf}
Generating electronic states over a dense enough $\bm{k}$-grid within the first Brillouin zone (BZ) is an essential step for producing continuous spectral functions that respect the continuity in the electronic density of states.
This is particularly important for simulating X-ray absorption spectra, especially when excited states extend into the continuum, either beyond the ionization potential in a non-infinite system or into the Bloch-periodic states of extended periodic systems. 
Although setting up a supercell for simulating XAS is equivalent to using some $\bm{k}$-point sampling over the BZ of the primitive unit cell, generating a $\bm{k}$-grid on the top of the supercell setups in some occasions does further improve the quality of simulated spectra, particularly at higher energies.
Similarly,  a metallic system may have a large number of extended states near the Fermi level, which need to be included to accurately reproduce the near-edge fine structure.
The number of extended states is proportional to and limited to the size of the supercell that can be realistically simulated in the $\Delta$SCF core-hole calculation.
In this circumstance, using $\bm{k}$-point sampling over the supercell BZ may partially compensate for the disadvantage of using a supercell that is not quite large enough. However,  $\bm{k}$-point sampling will not correct for a model of the final state within which the charge-density response to the core-excited state has not sufficiently converged within the supercell.

Previously, we have studied, implemented, and tested an efficient calculation scheme for obtaining band structure on a dense $\bm{k}$-grid. 
By employing so-called optimal basis sets \cite{shirley1996optimal, prendergast2009bloch},  one can first generate the band structure on a coarse $\bm{k}$-grid and then reproduce band energies and wave functions at any $\bm{k}$-point with much less computational effort.
The optimal basis set is the minimal basis for representing the periodic components of Bloch-waves across the BZ, constructed by removing linear dependence between these vectors (with the assumption that these functions vary smoothly throughout the BZ). 
Similar to plane-wave basis sets or maximally-localized Wannier functions \cite{marzari1997maximally}, the optimal basis functions, denoted as $\{B_i\}$, can be used to expand a Bloch-periodic wave function $|\psi_k\rangle=e^{ik \cdot r}|u_k\rangle$, in terms of its periodic component: $\langle B_i | u_k \rangle$, but are not limited to extended or localized states, no more than the actual Kohn-Sham orbitals themselves.  
Moreover, the number of optimal basis functions required is much smaller than the number of plane-waves for expanding these orbitals.
The size of a good optimal basis set ranges from $10^3$ to $10^4$, which can be at easily $1000$ times smaller than the plane-wave basis set of equivalent accuracy. The energies and eigenstates at a given k-point are obtained from diagonalization of a representation of the original (k-dependent) Kohn-Sham Hamiltonian in this much smaller basis.

Now we revisit the quantities needed for computing the overlap matrix elements in Eq. (\ref{eq:psipsi_final}), 
$         \langle \psi^\text{PS} | \tilde{\psi}^\text{PS} \rangle   $ and $       \langle p^I_{l} | \psi^\text{PS} \rangle (\langle \tilde{p}^I_{l} | \tilde{\psi}^\text{PS} \rangle)          $, 
which will benefit greatly from using optimal basis sets.
First, the pseudo overlap matrix element (carried out at every k-point independently) can be computed as
\begin{align}
\begin{split}
\langle \psi^\text{PS}_{nk} | \tilde{\psi}^\text{PS}_{mk} \rangle  = \sum_{ij} \langle u^\text{PS}_{nk} | B_i \rangle \langle B_i | \tilde{B}_j \rangle \langle \tilde{B}_j | \tilde{u}^\text{PS}_{mk} \rangle 
\end{split}
\end{align}
where $\langle u^\text{PS}_{nk} | B_i \rangle$ ($\langle \tilde{B}_j | \tilde{u}^\text{PS}_{mk} \rangle$) are the eigensolutions (Hermitian conjugates) of the k-dependent Hamiltonian in their corresponding optimal bases. Although each optimal basis set is constructed to be orthonormal, $\langle B_i | B_j \rangle = \delta_{ij}$ and $\langle \tilde{B}_i | \tilde{B}_j \rangle = \delta_{ij}$, note that the k-independent overlap matrix is not, in general: $\langle B_i | \tilde{B}_j \rangle \neq \delta_{ij}$, because we employ different optimal basis sets to represent initial- and final-state systems. We could in principle employ a sub-optimal basis to describe both systems, but it has not been attempted here. 

Although optimal basis functions themselves are represented in a plane-wave basis, $\{G_i\}$,  the relatively expensive calculation, 
$\langle B_i | \tilde{B}_j \rangle = \sum_{i'} \langle B_i | G_{i'} \rangle\langle G_{i'} | \tilde{B}_j \rangle $,
only needs to be computed once, and the matrix is universally applicable to any $\bm{k}$-point.
Similarly, $ \langle p^I_{l} | \psi^\text{PS} \rangle $ can be obtained by inserting the optimal basis set, in the same manner used to construct the same projectors in the non-local pseudopotential within the Hamiltonian.
This procedure has been implemented in the one-body core-hole approach and it simply needs to be extended to all atoms in the system.

\section{Spectral Convolution Theorem}
\label{sec:conv}
In practice, we may encounter a situation where a many-electron system can be factorizable into subsystems that are not entangled with each other, and inter-system transitions are forbidden.
For example, in a system where electron spins are collinear, and each electron can be associated with either a spin-up or -down state, then a many-body transition operator which can be similarly partitioned cannot induce transitions from the spin-up subsystem to the spin-down subsystem. The many-body dipole operator, which is the sum of one-body dipole operators, behaves in this way, and so, light-induced transitions of spin-collinear systems (within the dipole approximation) cannot effect spin cross-over. 

In general, if a spectrum reflects a multidimensional integral over a function factorizable for each independent variable (or, equivalently, over some partitioning of the same space), then we can take advantage of the spectral convolution theorem. For two independent variables, $x$ sampling subsystem $A$ and $y$ sampling subsystem $B$, suppose $f(x,y)=f_A(x) f_B(y)$, and we define a spectral function
\begin{align}
\begin{split}
\sigma(E) & = \int \int f(x,y) \delta(E-(x+y)) dx dy \\
& = \int \left( \int f_A(x) \delta((E-y)-x) dx \right) f_B(y) dy \\
& = \int \sigma_A(E-y) f_B(y) dy \\
& = \int \sigma_A(E-E') \sigma_B(E') dE'
\label{eq:specconv}
\end{split}
\end{align}
where we have just changed variable ($E'=y$) in the last line and defined the following subsystem spectral functions for each subset $I$ in the partition $\{A, B\}$:
\begin{align}
\begin{split}
\sigma_I(E)=\int_I f_I(E') \delta(E-E')
\end{split}
\end{align}
The general case, for many subsystems $I$ in $\{I_n\}$ can be written as a set of nested integrals over each subsystem,
\begin{align}
\begin{split}
\sigma(E)= & \int dE_1 \sigma_{I_1}(E-E_1) \\
                  & \times \int dE_2 \sigma_{I_2}(E_1-E_2) \\
                  & \dots \times \int dE_n \sigma_{I_{n-1}}(E_{n-1}-E_n) \sigma_{I_n} (E_n)  
\label{eq:nestedconv}
\end{split}
\end{align}

Let us assume that the many-body wave functions are factorizable and limit our discussion to two subsystems, $A$ and $B$, so that $|\Psi\rangle = |\Psi^A\rangle \otimes |\Psi^B\rangle$.
Then the transition amplitude can be factorized by considering final states where the transition probes each subsystem at a time, assuming that the transition operator can also be partitioned, for example, $\mathcal{O} = \sum_{i \in A} \mathcal{O}_i + \sum_{j \in B} \mathcal{O}_j$. Note that, in practice, if the symmetry of the system causes final states in different subsystems to be distinct, we should do a separate $\Delta$SCF calculation to define each final state orbital subspace. Here, let us focus on the components of the transition operator which act directly on subsystem $A$, inducing a many-body response in subsystem $B$, and index each final state by similarly partitioning the orbital configuration vector: $f = (f^A, f^B)$ (same for $i$) as follows: 
\begin{align}
\begin{split}
\langle \Psi_f |\mathcal{O}|\Psi_i\rangle = \langle \Psi_{f^A} |\mathcal{O}|\Psi_{i^A}\rangle
\langle \Psi_{f^B} | \Psi_{i^B}\rangle
\end{split}
\end{align}
Then the total spectrum can be written using Eq.~\ref{eq:specconv}, but recognizing a subtle difference between the subset spectral functions:
\begin{align}
\begin{split}
\sigma^A(E) &= \sum_{f_A} |\langle \Psi^A_{f_A} |\mathcal{O}|\Psi^A_i\rangle|^2 \delta(E - \Delta E_{f_A})
\end{split}
\end{align}
this includes the transition operator, while 
\begin{align}
\begin{split}
\sigma^B(E) &= \sum_{f_B} |\langle \Psi^B_{f_B} | \Psi^B_i\rangle|^2 \delta(E - \Delta E_{f_B})
\end{split}
\end{align}
reflects the response of subsystem $B$ to the excitation in $A$.
$\Delta E_f = \Delta E_{f_A} + \Delta E_{f_B}$ is the energy required to make the transition.

This theorem is particularly useful for combining spectra from opposite spin orientations and different k-points by performing each calculation separately.
In the many-electron formalism, the size of determinants for each subsystem is much smaller than the determinants for the entire system with spins taken into account, and hence one can compute a spectrum for each subsystem at much lower memory cost and time complexity and then obtain the resulting total spectrum via the nested spectral convolution outlined in Eq.~\ref{eq:nestedconv}.

\end{appendix}

%\bibliography{ref}{}
%\bibliographystyle{unsrt}

\end{document}